\documentclass[12pt,a4paper]{article} 

\usepackage[T1]{fontenc}
\usepackage[utf8]{inputenc}
\usepackage[USenglish]{babel}              
\usepackage[strict,autostyle]{csquotes}    

\usepackage{microtype}                     
\usepackage{setspace}                      
\usepackage{geometry}                      
\usepackage{titlesec}                      

\usepackage{amsmath}                       
\usepackage{amssymb}                       
\usepackage{amsthm}                        
\usepackage{enumitem}

\usepackage{graphicx}                      
\usepackage{booktabs}                      
\usepackage{tabularx}                      
\usepackage{multirow}                      
\usepackage{float}                         
\usepackage{rotating}                      
\usepackage{caption}                       
\usepackage{subcaption}                    

\usepackage{color}                         
\usepackage{xkcdcolors}                    
\usepackage[final]{draftwatermark}         

\usepackage{natbib}                        
\usepackage{hyperref}                      
\usepackage{url}                           

\usepackage{abstract}                      
\usepackage{authblk}                       
\usepackage{endnotes}                      
\usepackage{appendix}                      
\usepackage{multicol}                      
\usepackage{cuted}                         

\makeatletter
\renewcommand\tiny{\@setfontsize\tiny{5pt}{6pt}}
\renewcommand\scriptsize{\@setfontsize\scriptsize{7pt}{8pt}}
\renewcommand\footnotesize{\@setfontsize\footnotesize{8pt}{9.5pt}}
\renewcommand\small{\@setfontsize\small{9pt}{11pt}}
\renewcommand\normalsize{\@setfontsize\normalsize{10pt}{12pt}}
\renewcommand\large{\@setfontsize\large{12pt}{14pt}}
\renewcommand\Large{\@setfontsize\Large{14.4pt}{18pt}}
\renewcommand\LARGE{\@setfontsize\LARGE{17.28pt}{22pt}}
\renewcommand\huge{\@setfontsize\huge{20.74pt}{25pt}}
\renewcommand\Huge{\@setfontsize\Huge{24.88pt}{30pt}}
\makeatother

\usepackage{datetime}                      
\usepackage{etoolbox}                      
\RequirePackage[l2tabu,orthodox]{nag}      

\renewcommand{\arraystretch}{1}           

\titleformat{\section}{\normalfont\sffamily\large\bfseries\color{black}}{\thesection}{1em}{}
\titleformat{\subsection}{\normalfont\sffamily\bfseries\color{black}}{\thesubsection}{1em}{}

\hypersetup{
    pdffitwindow=true,                    
    breaklinks=true,                      
    colorlinks=false,                     
    pdfborder={0 0 0}                     
}

\newdateformat{monthyeardate}{\monthname[\THEMONTH] \THEYEAR}

\definecolor{darkgreen}{rgb}{0.0, 0.5, 0.0}
\definecolor{darkblue}{rgb}{0.0, 0.0, 0.55}
\definecolor{pink}{rgb}{1, 0.3, 0.6}
\definecolor{darkred}{rgb}{0.55, 0.0, 0.0}
\definecolor{brown}{rgb}{0.65, 0.16, 0.16}

\newcommand{\lab}[0]{\ensuremath{\ell}}
\newcommand{\diff}[0]{\ensuremath{\phi}}
\newcommand{\grain}[0]{\ensuremath{\mathrm{g}}}
\newcommand{\sym}[0]{\ensuremath{\mathrm{sym}}}

\newcommand{\equalcont}{\textsuperscript{\dag}}

\renewenvironment{abstract}
  {\normalsize
   \begin{center}
   \bfseries \sffamily \abstractname\vspace{-1em}\vspace{0pt}
   \end{center}
   \list{}{%
     \setlength{\leftmargin}{0.5cm}%
     \setlength{\rightmargin}{\leftmargin}%
   }%
   \item\relax}
  {\endlist}

\newenvironment{myappendix}{%
    \appendix
    \renewcommand{\thesection}{Appendix - \arabic{section}}
    \renewcommand{\thesubsection}{\thesection.\arabic{subsection}}
    \setcounter{section}{0}
    \setcounter{subsection}{0}
    \section*{Appendix}
    \addcontentsline{toc}{section}{Appendix}
    \preto\section{\ifnum\value{section}=0 \else\fi}
}{%
}

\author[1,2,3,\equalcont]{Brinthan Kanesalingam}
\author[1,2,4,\equalcont]{Darshan Chalise}
\author[5]{Carsten Detlefs}
\author[1,2,3,4,*]{Leora Dresselhaus-Marais}

\affil[1]{\small Materials Science and Engineering, Stanford University, Stanford, CA 94305, USA}
\affil[2]{\small SLAC National Accelerator Laboratory, Menlo Park, CA 94025, USA}
\affil[3]{\small PULSE Institute, Stanford University, Stanford, CA 94305, USA}
\affil[4]{\small SIMES, Stanford University, Stanford, CA 94305, USA}
\affil[5]{\small European Synchrotron Radiation Facility, 38043 Grenoble, France}

\affil[ ]{\small$^{\dagger}$These authors contributed equally to this work.}

\affil[ ]{\small$^{\ast}$Correspondence Email: \href{mailto:leoradm@stanford.edu}{leoradm@stanford.edu}}

\date{}

\title{\sffamily \textbf{Computation and Sensitivity Analysis of the Deformation-Gradient Tensor Reconstruction in Dark-Field X-ray Microscopy}}

\begin{document}
\maketitle
\vspace{-3em}
\begin{abstract}
    Spatially resolved strain measurements are crucial to understanding the properties of engineering materials. Although strain measurements utilizing techniques such as transmission electron microscopy (TEM) and electron backscatter diffraction (EBSD) offer high spatial resolution, they are limited to surface or thin samples. X-ray diffraction methods, including Bragg Coherent Diffraction Imaging and X-ray topography, enable strain measurements deep inside bulk materials but face challenges in simultaneously achieving both high spatial resolution and large field-of-view. Dark-field X-ray Microscopy (DFXM) offers a promising solution with its ability to image bulk crystals at the nanoscale while offering a field-of-view approaching a few hundred $\mu$m. However, an inverse modeling framework to explicitly relate the angular shifts in DFXM to the strain and lattice rotation tensors is lacking. In this paper, we develop such an inverse modeling formalism. Using the oblique diffraction geometry for DFXM, enabling access to noncoplanar symmetry-equivalent reflections, we demonstrate that the reconstruction of the full deformation gradient tensor ($\mathbf{F^{(g)}}$), and consequently the strain and lattice rotation matrices, is possible. In addition to the formalism, we also develop the computational framework to both forward calculate the anticipated angular shifts and reconstruct the average $\mathbf{F^{(g)}}$ for an individual pixel from DFXM experiments. Finally, utilizing the established formalism and computational framework, we present methods for sensitivity analysis to relate individual components of the rotation or strain tensor to specific angles of DFXM. The developed sensitivity analysis also enables explicit computation of the errors associated with the reconstruction of each component. The formalism, the computational framework, and the sensitivity analysis established in this paper should assist both the interpretation of past DFXM experiments and the design of future DFXM experiments. 
    
\end{abstract}
\newpage
\section{Introduction}
\label{sec:introduction}

Spatially resolved strain measurements play a critical role in understanding the behavior of dislocations and their influence on material properties, enabling quantitative analysis of local stress fields, dislocation densities, and strain gradients that govern plastic deformation mechanisms \citep{barabash2014strain}. Strain measurements in metals are essential to describe the loading states and dislocation densities relevant to mechanical properties and deformations \citep{cseren2023representative}. In functional materials, strain measurements elucidate mechanisms such as dendrite growth in solid-state lithium-ion batteries \citep{yildirim2024understanding}, or tracking thermally induced warpage in semiconductor packaging \citep{chalise2022temperature,tanner2021x}. Transmission Electron Microscopy \cite[TEM, ][]{hytch2014observing} and Electron Backscatter Diffraction \cite[EBSD, ][]{wilkinson2006high} are widely used for strain measurement due to their high spatial resolution and ability to provide detailed structural information. However, these methods are inherently limited to thin samples or surface regions. This restriction can modify the observed strain fields or change the dynamics predicted due to surface effects, making them unsuitable for analyzing strain deep within bulk materials.

Quantitative, spatially resolved strain measurements within bulk materials require noninvasive techniques capable of probing deep inside crystals. X-rays, with their high penetration depth compared to electron diffraction, enable imaging within bulk samples.  X-ray diffraction offers an opportunity to measure strain fields as any strain that results in the change in the spacing between lattice planes, $\Delta d_{hkl}$, affects the Bragg condition \citep{lester1925behavior}. Strain measurements along specific lattice vectors have been effectively conducted using monochromatic X-ray diffraction \cite[XRD, ][]{tanner2021x}. Measurement of every individual component of the 9-dimensional strain tensor, which quantifies all the longitudinal and shear strains along the three spatial directions, is generally more challenging. Techniques like wavelength-resolved Laue diffraction have established a formalism for measuring the full strain tensor using XRD \citep{liu2004three}. Spatially resolved strain measurements have become possible with techniques like scanning nano-diffraction, which extends non-spatially resolved methods by mapping strain distributions at the nanoscale \citep{meduvna2018lattice} . For higher defect-density crystals, Bragg Coherent Diffraction Imaging (BCDI) has been advanced with multi-reflection scanning and analysis, allowing spatially resolved  measurement of the strain tensor \citep{hofmann20173d}. However, BCDI generally provides a field-of-view limited to a few micrometers \citep{pateras2020combining}. Conversely, X-ray topography has been used for measurement of strain with a larger field-of-view \citep{tanner2013x}. However, X-ray topography is usually limited to measurement of axial strain with spatial resolution limited to $\sim$10~$\mu$m \citep{tanner2013x}. Characterizing strain fields and deformations at the mesoscale in bulk crystals-requiring spatial resolution approaching 100~nm and a field of view of about 100~$\mu$m—remains a significant challenge.

Dark-field X-ray microscopy (DFXM) is a full-field X-ray imaging technique that employs an objective lens aligned with a diffracted X-ray beam to produce real-space images of a crystal \citep{simons2015dark}. This method enable depth-resolved imaging of samples with spatial resolution reaching approximately 100~nm \citep{carlsen2022fourier}. The objective lens, often a compound refractive lens (CRL), projects a magnified image of the sample onto a detector \citep{simons2017simulating}. Furthermore, the objective lens  is an aperture in reciprocal space, achieving a resolution reaching more than 10$^{-4}$ radians \citep{poulsen2017x} when used with monochromatic and nearly collimated X-rays generated by advanced 4th-generation synchrotron and X-ray free-electron lasers (XFEL) sources. This high resolution in reciprocal space enables detailed measurements of shear strain and lattice rotations. With superior resolution in reciprocal space and the ability to image 3D bulk crystals using high-energy X-rays, DFXM is highly effective for visualizing localized strain fields and dislocation structures deep within bulk materials \citep{dresselhaus2021situ, yildirim2023extensive, pal2025measuring}.

Formalisms developed by \cite{poulsen2017x,poulsen2021geometrical} have enabled forward modeling of DFXM images at specific goniometer angles and CRL positions. These approaches rely on specifying the deformation gradient tensor $\mathbf{F^{(g)}}$, comprising lattice rotation ($\mathbf{w}$) and strain ($\mathbf{\varepsilon}$), at each grid point in the simulation\footnote{To be consistent with the notation used in \cite{detlefs2025oblique}, we use the notation $\mathbf{w}$ for the lattice rotation tensor and $\omega$ for the goniometer rotation angle.}. The formalisms calculate the normalized difference between the X-ray scattering vectors for undeformed and deformed samples in the imaging coordinate system, defined by goniometer motor positions, CRL positions, and scattering geometry \citep{poulsen2021geometrical}. Image contrast is determined using this normalized difference vector and the reciprocal space resolution, providing insights into experimental DFXM images. This method was used by \cite{borgi2024simulations} to simulate the contrast of isolated dislocations. However, an inverse modeling formalism to directly relate angular shifts in DFXM quantified by goniometer and CRL/detector motor positions at maximum pixel intensity—to the explicit components of strain and lattice rotation tensors remains undeveloped. Establishing such a framework would significantly enhance the quantitative capabilities of DFXM for strain and lattice rotation measurements, further advancing its application in materials science.

\section{Established Modeling Frameworks}

We begin discussing our approach to model the full $\mathbf{F^{(g)}}$ by establishing the current frameworks used to model DFXM image contrast and the micromechanical models that are used to describe deformations in a lattice. In Section~\ref{sec:review_of_dfxm_formalism}, we introduce the formalisms described in \cite{poulsen2017x,poulsen2021geometrical,simons2017simulating} in which an assumption of kinematic diffraction enables the authors to devise an analytical model for how incident beam parameters and instrumentation give rise to diffracted signals for a given lattice system. We then mention how the formalisms define an arbitrary deformation via a micromechanical model to describe how a known deformation gives rise to specific DFXM signals and associated image contrast. In Section~\ref{micromech}, we then introduce the micromechanical modeling framework first introduced in \cite{poulsen2021geometrical}, and extend that framework to a more general solution to describe a $\mathbf{F^{(g)}}$ that will form the basis for the present study.

\subsection{Review of DFXM Formalism}
\label{sec:review_of_dfxm_formalism}

In this section, we introduce the components of the DFXM setup to establish the basis for the diffraction formalism introduced in \cite{poulsen2021geometrical}. To do this, we describe the different components of the DFXM experiment, starting by defining a series of coordinate systems that relate the different components of the DFXM instrument and their spatial relationship. We then describe the parameters of the incident and diffracted beams and finally introduce the sample orientation modulated by the angles of the goniometer. We conclude with a brief introduction of the work that has recently begun to demonstrate the geometry to achieve full $\mathbf{F^{(g)}}$ in DFXM.

\begin{figure}[!t]
    \centering    \includegraphics[width=0.7\linewidth]{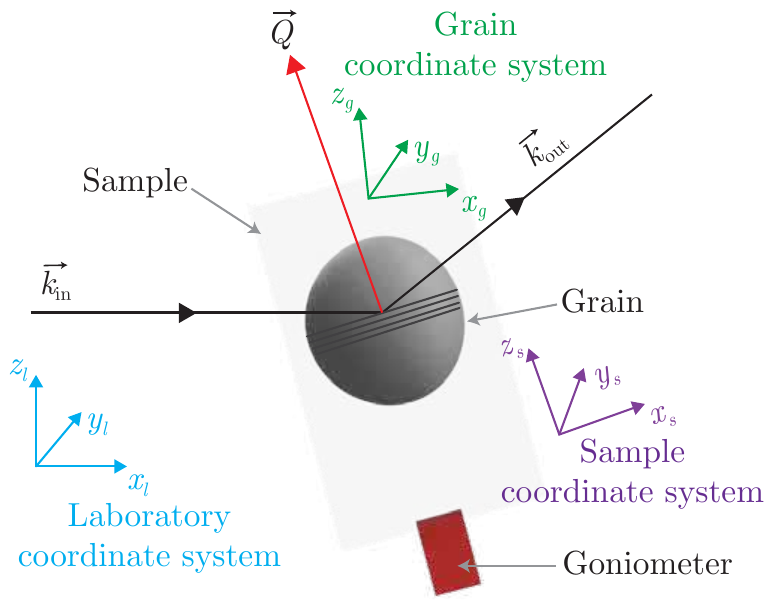}
    \caption{Coordinate systems in DFXM.}
    \label{fig:coordinate_systems}
\end{figure}

The DFXM instrument includes several different components whose relative positions and motions must be described in relation to each other. For mathematical ease, we define a coordinate system to describe each component's position independently, with transformation matrices describing the relative positions in an easily adjustable manner. All coordinate systems have a set of characteristic vectors in real and reciprocal space. The direction of the incident X-ray beam defines the laboratory coordinate system. The conventions for the directions, $\hat{x}_{\ell}$ sets along the direction of the incident wavevector $\vec{k}_{\text{in}}$, $\hat{z}_{\ell}$ vertically upward, and the $\hat{y}_{\ell}$ by the right-hand rule. The orientation of the sample defines the sample coordinate system, \{$\hat{x}_s$,$\hat{y}_s$,$\hat{z}_s$\}, which relates to the lab coordinate system based on the sample's orientation that is set by the goniometer stages' rotational positions (described in $\mathbf{\Gamma}$). Other works describe the sample coordinate system as the goniometer or diffractometer coordinate system \citep{detlefs2025oblique}. The crystallographic orientation of the grain involved in diffraction defines the grain coordinate system \{$\hat{x}_g$, $\hat{y}_g$, $\hat{z}_g$\}. This system is often defined by the basis vectors (1 0 0), (0 1 0), and (0 0 1) directions in Miller-index notation for the grain of interest. Position vectors and diffraction vectors in the sample and grain coordinate systems relate through the orientation matrix ($\mathbf{U}$). Figure~\ref{fig:coordinate_systems} demonstrates the previously discussed coordinate systems within DFXM. 
\begin{equation}
    \begin{array}{ccccccc}
    \text{Miller Indices} 
    & \xrightarrow{\mathbf{B}}
    & \text{Grain c.s.}
    & \xrightarrow{\mathbf{U}}
    & \text{Diffractometer c.s.}
    & \xrightarrow{\mathbf{\Gamma}}
    & \text{Laboratory c.s.}
    \\
    (h, k, \ell) 
    & 
    & (x_g, y_g, z_g)
    & 
    & (x_s, y_s, z_s)
    & 
    & (x_\ell, y_\ell, z_\ell)
    \end{array}
\end{equation}
Based on the above coordinate systems and the wavevectors of the incident and diffracted beams, we now describe the geometry of the DFXM instrument. DFXM experiments uses nearly monochromatic and collimated X-rays. The wavevector of the beam that illuminates the sample, $\vec{k}_{\text{in}}$, has a magnitude that is related to the selected beam energy, $E$. The diffracted wave vector is denoted by $\vec{k}_{\text{out}}$, and the corresponding scattering vector is defined as $\vec{Q} = \vec{k}_{\text{out}}-\vec{k}_{\text{in}}$. For elastic scattering, $\left| \vec{k}_{\text{in}} \right| = \left| \vec{k}_{\text{out}} \right| = k = 2\pi/\lambda$, where $\lambda$ is the X-ray wavelength.
The Bragg condition is fulfilled when the scattering vector $\vec{Q}$ is equal a reciprocal lattice vector $\vec{G}$ associated with the Miller indices $(h,k,\ell)$ and given by
\begin{equation}
    \vec{Q}_\ell = \vec{G}_\ell = \mathbf{\Gamma} \mathbf{U} \mathbf{B} \left( \begin{array}{c} h \\ k \\ \ell \end{array} \right).
\end{equation}

The incident beam illuminates the sample, which is mounted on a goniometer whose base tilt, denoted by $\mu$, is defined with respect to the direction of the $\vec{k}_\text{in}$. The position of the goniometer determines the orientation of the sample with respect to the incident beam, and thus the tilt of the $hk\ell$ plane required to satisfy the Bragg condition. The orientation of the sample can be adjusted using goniometer angles, $\omega, \phi$, and $\chi$. The diffracted beam defines the position of the detector and CRL. The detector and CRL align collinearly with $\vec{k}_\text{out}$ by moving the detector and CRL collectively using $2\theta$ (scattering angle) and $\eta$ (azimuthal angle) angles. 

In the simplified geometry described by \cite{poulsen2021geometrical}, a crystallographic plane of the form (00$\ell$) is chosen for diffraction, and the goniometer is tilted by the Bragg angle ($\mu = \theta_B$) to fulfill the Bragg condition. This configuration simplifies the DFXM setup: the scattering angle becomes $2\theta_B$, and the azimuthal angle is fixed at $\eta = 0$. Under this specific alignment, the orientation matrix $\mathbf{U}$ reduces to the identity matrix, indicating no misorientation between the grain and sample coordinate systems. Consequently, the goniometer rotation $\omega$ corresponds to a pure rotation about the diffraction vector $\vec{Q}$, as shown in Figure~\ref{fig:oblique_diffraction_geometry_2}(a).

\subsection{Oblique Geometry for Multiple Reflections}

\begin{figure}[!t]
    \centering
\includegraphics[width=\linewidth]{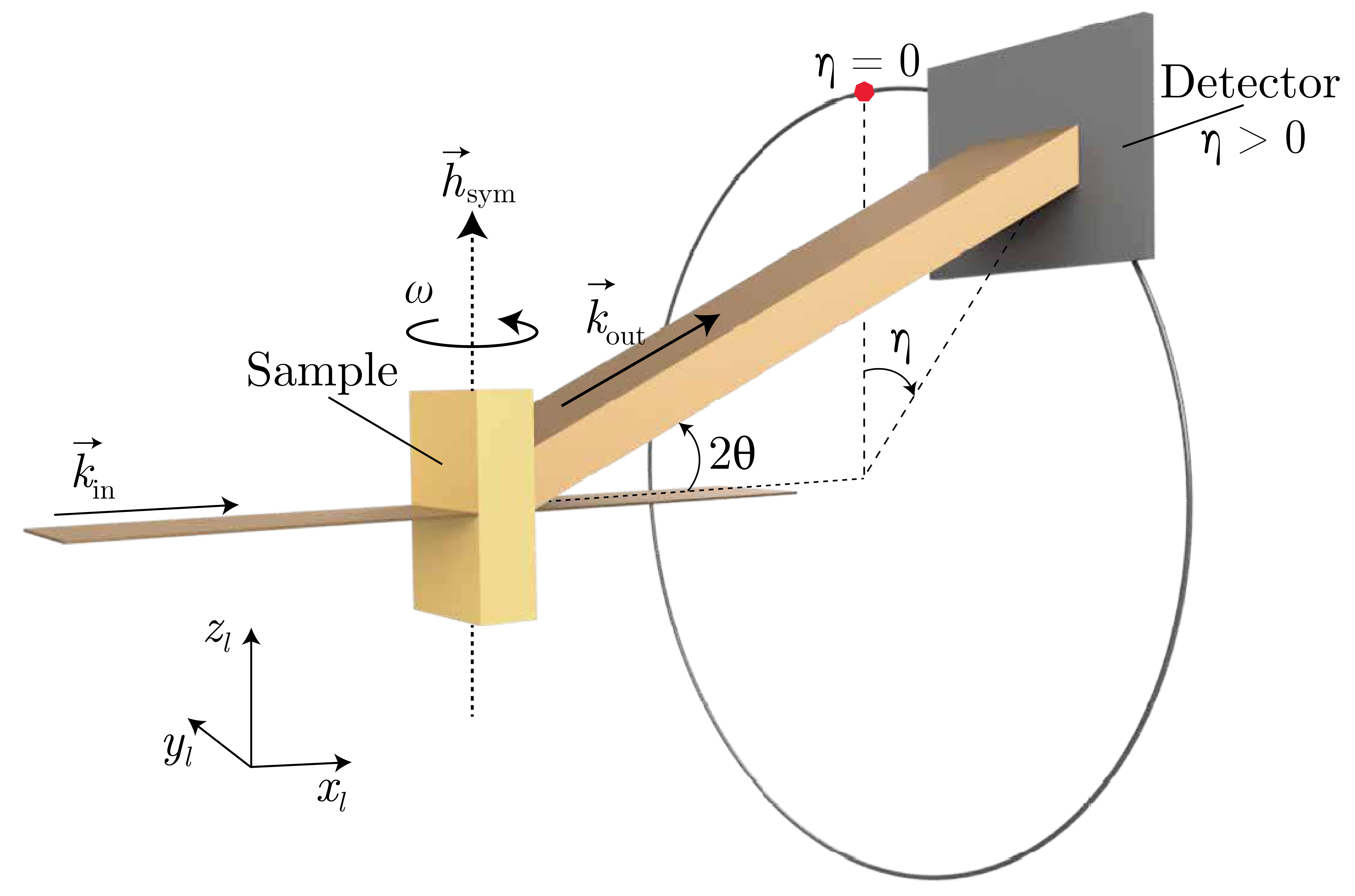}
    \caption{Oblique diffraction geometry in DFXM. Vertical diffraction geometry for multiple reflections without changing the illumination volume. Image adapted and modified from \cite{detlefs2025oblique}.}
\label{fig:oblique_diffraction_geometry}
\end{figure}

\begin{figure}[!t]
    \centering
\includegraphics[width=\linewidth]{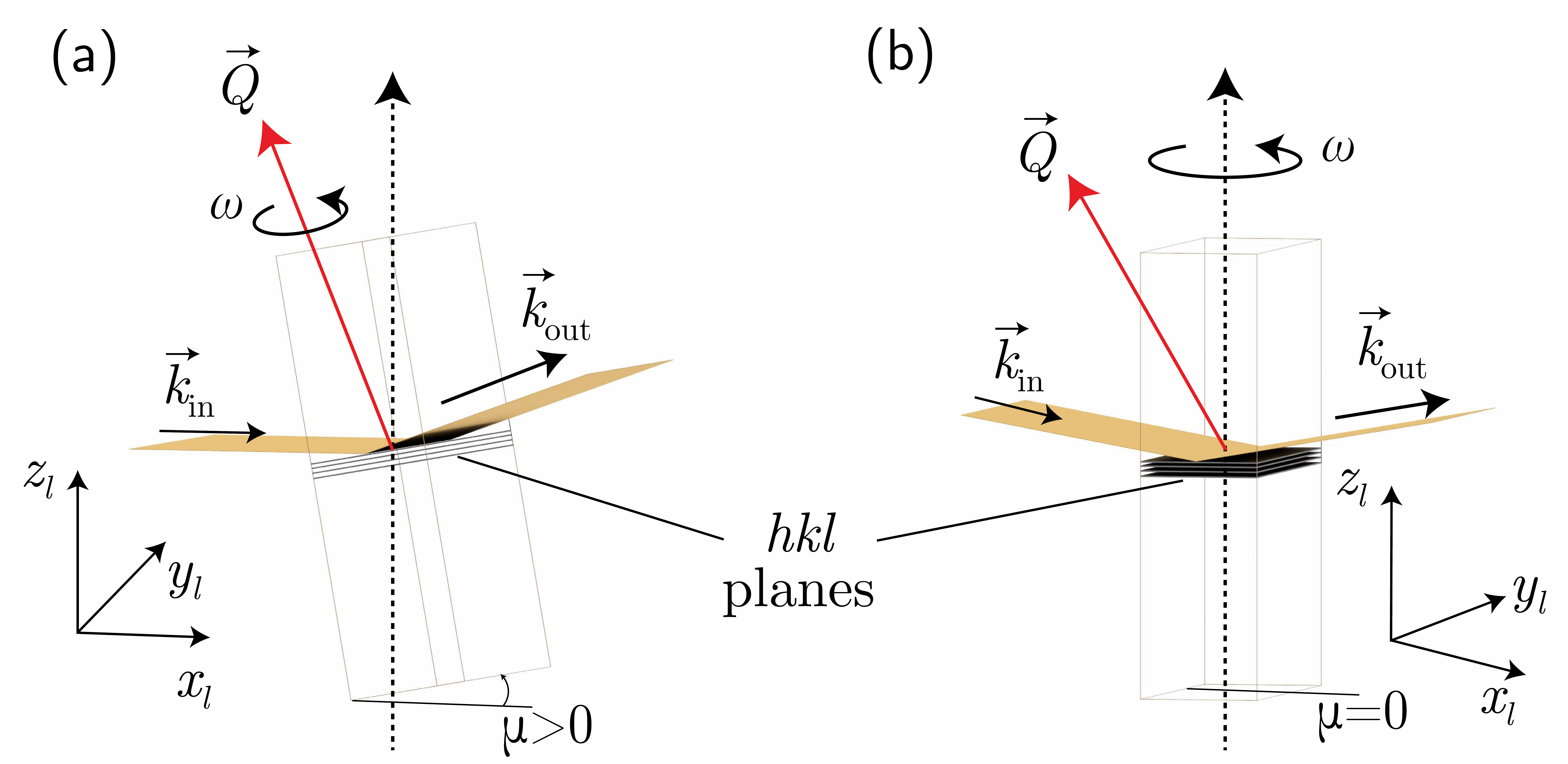}
    \caption{Simple versus oblique diffraction geometry in DFXM. (a) Simplified geometry for diffraction where scattering vector ($\vec{Q}$) and goniometer rotation axis ($\omega$) are collinear (b) $\vec{Q}$ and $\omega$ are non-collinear for a reflection in oblique diffraction geometry.}
\label{fig:oblique_diffraction_geometry_2}
\end{figure}

In this study, we adopt the oblique diffraction geometry as proposed by \cite{detlefs2025oblique}. This geometry, illustrated in Figure~\ref{fig:oblique_diffraction_geometry}, differs from the simplified geometry described in \cite{poulsen2021geometrical}, where a nonzero base tilt ($\mu \neq 0$) aligns the diffraction vector $\vec{Q}$ with the goniometer rotation axis $\omega$. In contrast, the oblique geometry sets $\mu = 0$, resulting in a non-parallel relationship between $\vec{Q}$ and $\omega$. For a cubic crystal system, the angle between these two directions equals the Bragg angle $\theta_B$, as shown in Figure~\ref{fig:oblique_diffraction_geometry_2}(b). This geometry enables access to non-coplanar, symmetry-equivalent reflections. Such reflections are critical for resolving all components of the $\mathbf{F^{(g)}}$, as they offer linearly independent geometric projections—something not achievable with coplanar reflections alone. Additionally, symmetry-equivalent reflections can be accessed by varying only the $\omega$ rotation, while maintaining a constant azimuthal angle $\eta$ and preserving the illuminated volume. The resulting geometry defines an oblique diffraction condition, where the diffracted wavevector $\vec{k}_{\text{out}}$ is characterized by a scattering angle $2\theta$ and an azimuthal angle $\eta$. We refer readers to \cite{detlefs2025oblique} for further information of the oblique geometry. 

\subsection{Micromechanical Framework}
\label{micromech}

DFXM measurements are fundamentally connected to the measurements of the deformation gradient tensor $\mathbf{F^{(g)}}$ (refer to Eq.~\ref{eq:diffraction_condition_deformed}), which maps material vectors from a reference configuration to their current (deformed) state. The deformation gradient field can be represented by the product of a rotation matrix, $\mathbf{R}_1$, and the right stretch tensor, $\mathbf{V}_1$, or equivalently the product of the left stretch tensor $\mathbf{V}_2$ \citep{poulsen2021geometrical}, i.e.,
\begin{equation}
    \mathbf{F^{(g)}} =  \mathbf{R}_1 \mathbf{V}_1 = \mathbf{V}_2 \mathbf{R}_2.
    \label{eq:deformation_gradient}
\end{equation}

In the conventional DFXM formalism \citep{poulsen2021geometrical} an infinitesimal deformation approximation (linear approximation) is typically used to decompose the retrieved $\mathbf{F^{(g)}}$ to strain and lattice rotations. Within this approximation, the $\mathbf{F^{(g)}}$ can be represented as a combination of the strain tensor ($\mathbf{\varepsilon}$) and the  lattice rotation tensor ($\mathbf{w}$), i.e.,
\begin{equation}
    \mathbf{F^{(g)}} \approx \mathbf{\varepsilon} + \mathbf{w} + \mathbf{I}.
    \label{eq:deformation_gradient_tensor}
\end{equation}

The $\mathbf{\varepsilon}$ is the symmetric Biot strain tensor with six independent components and $\mathbf{w}$ is the antisymmetric lattice rotation tensor with three free parameters.

However, this linear approximation cannot be generalized to all experimental scenarios. In some experimental cases, there would be instances where deformations exceed the linear approximation range. Examples of cases exceeding the linear approximation  include shock wave compression experiments \citep{katagiri2023transonic, he2022diamond}, laser-induced deformation and heating \citep{jaisle2023mhz,ball2024measurement}, plastic deformations \citep{cretton2025observation,zelenika2025observing}, additive manufacturing and thermal processing \citep{liesegang2023investigation}, highly strained heterostructures \citep{chen20253d,hill20223d}, and complex dislocation structures \citep{wang2025computing}. For these finite deformation cases, a more rigorous formulation is required.

$\mathbf{F^{(g)}}$ is a fundamental two-point tensor that maps infinitesimal material vectors from the reference configuration to the current configuration. In the finite deformation theory, the decomposition of $\mathbf{F^{(g)}}$ expressed in Eq.~\ref{eq:deformation_gradient} remains. However, strain measurements require modifications to account for the nonlinear effects that become prominent in large deformation scenarios. In particular, the quadratic term $\mathbf{H}^T \mathbf{H}$—where $\mathbf{H}$ denotes the displacement gradient—is typically neglected under the assumption of small deformations but becomes non-negligible in finite deformation regimes. There, instead of using the infinitesimal Biot strain tensor $\mathbf{\varepsilon}$, finite deformation theory uses the Green-Lagrange strain tensor $\mathbf{E}$ \citep{holzapfel2002nonlinear, truesdell2004non}, defined as:
\begin{equation}
\mathbf{E} = \frac{1}{2}(\mathbf{C} - \mathbf{I})
\end{equation}
where $\mathbf{C}$ is the right Cauchy-Green deformation tensor given by:
\begin{equation}
\mathbf{C} = \mathbf{F^{(g)}}^T\mathbf{F^{(g)}} = \mathbf{V}_1^T \mathbf{R}^T \mathbf{R} \mathbf{V}_1 = \mathbf{V}_1^2
\end{equation}
The Green-Lagrange strain tensor properly accounts for the nonlinear effects of finite deformation, incorporating both the linear terms present in infinitesimal theory and the additional nonlinear terms that become significant in large deformation scenarios (A detailed derivation is explained in \ref{appendix-finitedeformation}).

The relationship between the infinitesimal strain tensor $\mathbf{\varepsilon}$ and the Green-Lagrange strain tensor $\mathbf{E}$ can be established by examining the limit of small deformations:
\begin{equation}
\lim_{\|\mathbf{F^{(g)}} - \mathbf{I}\| \to 0} \mathbf{E} = \mathbf{\varepsilon}
\label{eq:strain_limit}
\end{equation}
This relationship demonstrates that the infinitesimal strain tensor is a first-order approximation of the Green-Lagrange strain tensor for small deformations.

The finite lattice rotation tensor $\mathbf{w}$ is obtained via the matrix logarithm:
\begin{equation}
    \mathbf{w} = \log(\mathbf{R}),
\end{equation}
where $\mathbf{w}$ is skew-symmetric ($\mathbf{w}^\mathsf{T} = -\mathbf{w}$).

This Cauchy-Green deformation treatment provides a fully generalizable formulation of strain through $\mathbf{F^{(g)}}$. Importantly, this generalization does not preclude the measurement of $\mathbf{F^{(g)}}$ that came from previous derivations by \cite{poulsen2021geometrical,detlefs2025oblique}. Rather, it extends the applicability of the $\mathbf{F^{(g)}}$ framework to arbitrary systems of strain and lattice, capable of accurately describing finite deformations beyond the linear approximation regime.

\section[Reconstructing F from Goniometer and Detector Positions in DFXM]{Reconstructing $\mathbf{F^{(g)}}$ from Goniometer and Detector Positions in DFXM }
\label{Reconstructing the F from DFXM observables}

In this section, we demonstrate how all components of $\mathbf{F^{(g)}}$ can be reconstructed from DFXM measurements. We do this by determining the goniometer ($\phi$, $\chi$, $\mu$, and $\omega$) and detector positions ($\theta$ and $\eta$) resulting in the  maximum intensity at a given voxel for 3 non-coplanar symmetry-equivalent reflections. The equations accomplishing this derivation are presented in the sections below.

In Section~\ref{sec:general_formalism}, we establish the fundamental equations, relating the components of $\mathbf{F^{(g)}}$ to the goniometer and detector positions corresponding to the maximum intensity of a voxel for an individual reflection. We build on to these equations in Section~\ref{multireflection}, to relate the goniometer and detector positions to the components of $\mathbf{F^{(g)}}$ for three symmetry-equivalent reflections and provide equations that enable the full  reconstruction of $\mathbf{F^{(g)}}$. 

\subsection[Relating DFXM Goniometer and Detector Positions to F for a Single Reflection]{Relating DFXM Goniometer and Detector Positions to $\mathbf{F^{(g)}}$ for a Single Reflection}
\label{sec:general_formalism}
 
In DFXM experiments, we measure an image along $\vec{k}_{\text{out}}$ at a given crystal orientation and detector position. As we rotate the crystal, changing the orientation and associated regions that meet the Bragg condition, the image intensities change in a way that correlate the $\mathbf{F^{(g)}}$ as it is projected onto the diffraction vector $\vec{Q}$. In the sample coordinate system, the diffraction vector can be defined using the Laue condition as
\begin{equation}
    \vec{Q}_s = \mathbf{U} \mathbf{B} \left( \begin{array}{c} h \\ k \\ \ell \end{array} \right),
    \label{eq:diffraction_vector_sample}
\end{equation}
where $\mathbf{U}$ is the orientation matrix that converts the diffraction vector, $\vec{Q}$ from the grain coordinate system to the sample coordinate system \citep{busing1967angle}, $\mathbf{B}$ is the reciprocal space basis matrix of the crystal \citep{bernier2011far}, and \textit{hkl} are the Miller indices of the crystallographic reflection under consideration by the model  \citep{poulsen2017x,poulsen2021geometrical}. For an undeformed crystal, we represent an undeformed lattice by $\mathbf{B}_0$. We define the modification of $\mathbf{B}_0$ by any arbitrary deformation (for a deformed crystal) as
\begin{equation}
    \mathbf{B} = \left( \mathbf{F^{(g)}} \right)^{-T} \mathbf{B}_0
    \label{eq:deformed_B}.
\end{equation}

The conversion of $\vec{Q}$ from the sample to laboratory coordinate systems is given by
\begin{equation}
    \vec{Q}_\ell = \mathbf{\Gamma} \vec{Q}_s,
    \label{eq:diffraction_vector_lab}
\end{equation}
where $\mathbf{\Gamma}$ is the product of rotation matrices, each of which describe the rotation of certain goniometer axis, $\mathbf{\Gamma}$ given by
\begin{equation}
    \mathbf{\Gamma(\mu, \omega, \chi, \phi)} = \mathbf{M}(\mu) \mathbf{\Omega}(\omega) \mathbf{X}(\chi) \mathbf{\Phi}(\phi).
    \label{eq:combined_rotation_matrix_hennings}
\end{equation}
Here, $\phi$, $\chi$, $\omega$, and $\mu$ are the Euler angles for the right-handed rotation matrices. $\mu$ is the base tilt about the $y_{\ell}$ axis, $\omega$ is the rotation around the axis of goniometer, and $\chi$ and $\phi$ are two orthogonal motors that tilt the sample in perpendicular directions to the axis of $\omega$.

We note that the order of the rotation presented in the above equation may not necessarily be the same across all DFXM experiments. In the cases when the order of the rotation is different, the arrangement of $\mathbf{\Gamma}$ should be different to account for the order.

The order of rotation matrices following the convention according to the geometry of the experiments takes care of the proper alignment of diffraction vector from sample to laboratory coordinate system.

In \cite{poulsen2017x}, the form of $\vec{Q}_\ell$ that meets the Bragg condition may be derived from $\vec{k}_\text{in}$ and $\vec{k}_\text{out}$ based on the detector position as
\begin{equation}
    \vec{Q}_\ell = \frac{4\pi}{\lambda} \sin(\theta) \left( \begin{array}{ccc} -\sin(\theta) \\ -\cos(\theta) \sin(\eta) \\ \cos(\theta) \cos(\eta) \end{array} \right),
    \label{eq:scattering_vector_lab}
\end{equation}
where $\lambda$ is the wavelength of the X-ray beam. From \cite{poulsen2021geometrical}, we get
\begin{equation}
    \mathbf{\Gamma(\mu, \omega, \chi, \phi)} \vec{Q}_s = \frac{4\pi}{\lambda} \sin(\theta) \left( \begin{array}{ccc} -\sin(\theta) \\ -\cos(\theta) \sin(\eta) \\ \cos(\theta) \cos(\eta) \end{array} \right).
    \label{eq:diffraction_condition_general}
\end{equation}
While the above equations may be generalized to any diffraction geometry, the assignment of each goniometer angle to specific rotational axes in the lab coordinate system imposes a specific geometry to the experiment.

For a deformed crystal with $\mathbf{F^{(g)}}$, the Eq.~\ref{eq:diffraction_condition_general} will modify to
\begin{equation}
    \mathbf{M}(\mu) \mathbf{\Omega}(\omega)\mathbf{X}(\chi)\mathbf{\Phi}(\phi) \mathbf{U} (\mathbf{F^{(g)}})^{-T} \mathbf{B}_0 \left( \begin{array}{c} h \\ k \\ \ell \end{array} \right) = \frac{4\pi}{\lambda} \sin(\theta) \left( \begin{array}{ccc} -\sin(\theta) \\ -\cos(\theta) \sin(\eta) \\ \cos(\theta) \cos(\eta) \end{array} \right).
    \label{eq:diffraction_condition_deformed}
\end{equation}

Here, we assume $\mathbf{B}_0$ to be known from the nominal lattice constants \citep{poulsen2021geometrical}. The orientation of the crystal on the goniometer, $\mathbf{U}$, is obtained by measuring the goniometer angles of two non-collinear reflections \citep{busing1967angle}. For simulations, $\mathbf{U}$ can be chosen, e.g., following the procedure described in \ref{appendix1}. The purpose of this formalism is to determine $\mathbf{F^{(g)}}$ via $(\mathbf{F^{(g)}})^{-T}$.

For this formalism, to avoid taking the inverse of the $\mathbf{F^{(g)}}$, we simplify the formalism by defining
\begin{equation}
    \mathbf{T} = (\mathbf{F^{(g)}})^{-T}.
    \label{eq:deformation_gradient_inverse}
\end{equation}

Here $\mathbf{T}$ is a 3 × 3 matrix (also this is equivalent to the $(\mathbf{H} + \mathbf{I})$ in \cite{poulsen2021geometrical})\footnote{$\mathbf{T}$ is different from the left stretch tensor defined in the \citep{poulsen2021geometrical}}. 



Using Eq.~\ref{eq:deformation_gradient_inverse}, Eq.~\ref{eq:diffraction_condition_deformed} can be written as

\begin{align}
    \mathbf{\Gamma(\mu, \omega, \chi, \phi)} \mathbf{U} \mathbf{T} \mathbf{B}_0 \begin{pmatrix} h \\ k \\ \ell \end{pmatrix} =& \frac{4\pi}{\lambda} \sin(\theta) \begin{pmatrix} -\sin(\theta) \\ -\cos(\theta) \sin(\eta) \\ \cos(\theta) \cos(\eta) \end{pmatrix},
    \label{eq:diffraction_condition_deformed_final}\\
    \mathbf{T} \begin{pmatrix} B_{0_{11}} h + B_{0_{12}} k + B_{0_{13}} \ell \\ B_{0_{21}} h + B_{0_{22}} k + B_{0_{23}} \ell\\ B_{0_{31}} h + B_{0_{32}} k + B_{0_{33}} \ell \end{pmatrix} =& \frac{4\pi}{\lambda} (\mathbf{\Gamma(\mu, \omega, \chi, \phi)} \mathbf{U})^{-1} \begin{pmatrix} -\sin^2(\theta) \\ -\cos(\theta) \sin(\eta) \sin(\theta) \\ \cos(\theta) \cos(\eta) \sin(\theta) \end{pmatrix}.
    \label{eq:diffraction_condition_deformed_final_1}
\end{align}

Here, $B_{0ij}$ terms are the components of the $\mathbf{B}_0$ matrix and it could be defined in several ways including \cite{busing1967angle, Chung1999, Schlenker1978, poulsen2017x}. Then, Eq.~\ref{eq:diffraction_condition_deformed_final_1} can be written as,
\begin{equation}
    \mathbf{T} \left( \begin{array}{c} h^\prime \\ k^\prime \\ \ell^\prime \end{array} \right) = \mathbf{P}(\mu, \omega, \chi, \phi, \lambda)  \left( \begin{array}{ccc} -\sin(\theta)^2 \\ -\cos(\theta) \sin(\eta) \sin(\theta) \\ \cos(\theta) \cos(\eta) \sin(\theta) \end{array} \right)
    \label{eq:diffraction_condition_deformed_final_2}
\end{equation}
where $\mathbf{P} = (4\pi / \lambda) (\mathbf{\Gamma(\mu, \omega, \chi, \phi)} \mathbf{U})^{-1}$ and it is a function of $\mu, \omega, \chi, \phi, \text{and } \lambda$. $P_{ij}$ terms are the components of the $\mathbf{P}$ matrix. Further, let  $\left( \begin{array}{ccc} h^\prime & k^\prime & \ell^\prime \end{array} \right)^T$ represent the undeformed diffraction vector in the grain coordinate system, i.e., $\left( \begin{array}{ccc} h^\prime \\ k^\prime \\ \ell^\prime \end{array} \right)  =\begin{pmatrix} B_{0_{11}} h + B_{0_{12}} k + B_{0_{13}} \ell \\ B_{0_{21}} h + B_{0_{22}} k + B_{0_{23}} \ell\\ B_{0_{31}} h + B_{0_{32}} k + B_{0_{33}} \ell \end{pmatrix}$. Then, Eq.~\ref{eq:diffraction_condition_deformed_final_2} can be written as a system of three equations
\begin{equation}
    \resizebox{\textwidth}{!}{$
    \left\{
    \begin{aligned}
        h^\prime T_{11} + k^\prime T_{12} + \ell^\prime T_{13} = -P_{11}(\mu, \omega, \chi, \phi, \lambda) \sin(\theta)^2 - P_{12}(\mu, \omega, \chi, \phi, \lambda) \cos(\theta) \sin(\eta) \sin(\theta) + P_{13}(\mu, \omega, \chi, \phi, \lambda) \cos(\theta) \cos(\eta) \sin(\theta) \\
        h^\prime T_{21} + k^\prime T_{22} + \ell^\prime T_{23} = -P_{21}(\mu, \omega, \chi, \phi, \lambda) \sin(\theta)^2 - P_{22}(\mu, \omega, \chi, \phi, \lambda) \cos(\theta) \sin(\eta) \sin(\theta) + P_{23}(\mu, \omega, \chi, \phi, \lambda) \cos(\theta) \cos(\eta) \sin(\theta) \\
        h^\prime T_{31} + k^\prime T_{32} + \ell^\prime T_{33} = -P_{31}(\mu, \omega, \chi, \phi, \lambda) \sin(\theta)^2 - P_{32}(\mu, \omega, \chi, \phi, \lambda) \cos(\theta) \sin(\eta) \sin(\theta) + P_{33}(\mu, \omega, \chi, \phi, \lambda) \cos(\theta) \cos(\eta) \sin(\theta).
    \end{aligned}
    \right.
    $}
    \label{eq:diffraction_equation_full_form_single_reflection_h}
\end{equation}

The resulting Eq.~\ref{eq:diffraction_equation_full_form_single_reflection_h} is a system of three equations with nine unknowns (the nine components of $\mathbf{F^{(g)}}$). 

Eq.~\ref{eq:diffraction_equation_full_form_single_reflection_h} is general across any reflection probed via DFXM. This equation itself is comprehensive enough to calculate the angular shifts caused by $\mathbf{F^{(g)}}$ on a single reflection. Therefore,  the formalism provided in this subsection is valuable to the DFXM community in planning the experiment interpreting the DFXM measurements when single reflections are measured.

In cases where only one component of $F_{ij}$ is unknown for every $i$'th row vector in $\mathbf{F^{(g)}}$, full reconstruction of $\mathbf{F^{(g)}}$ using Eq.~\ref{eq:diffraction_equation_full_form_single_reflection_h} is possible. 
In \ref{sec:strain_wave}, we present the analytical reconstruction in such a case - an example of the deformation caused by a longitudinal strain wave along the [100] direction in diamond. 

\subsection[Complete Reconstruction of F from Multiple Reflections]{Complete Reconstruction of $\mathbf{F^{(g)}}$ from Multiple Reflections}
\label{multireflection}

In scenarios where there are no restraints on any component of the $\mathbf{F^{(g)}}$, the reconstruction of $\mathbf{F^{(g)}}$ would require at least nine independent equations corresponding to three non-coplanar reflections \citep{detlefs2025oblique}. We build the formalism for such reconstruction in this section.

If there are spatial variations of the strain fields throughout the volume of the crystal, then registering the spatially resolved measurements of three or more reflections into a volume model of the sample is a challenge. One approach to facilitate this is to choose an experimental configuration that keeps the illumination volume constant for all measured reflections. Such a geometry has been proposed by \cite{detlefs2025oblique} (we also note that such geometry has been utilized by \cite{Chung1999} and \cite{Abboud2017} for white beam Laue diffraction). In brief, \cite{detlefs2025oblique} achieved this by the rotation around the $\omega$ ($z_{l}$ axis or $\vec{h}_{sym}$) without changing the orientation of sample in any other axes without changing the illumination volume in the sample as shown in the Figure~\ref{fig:oblique_diffraction_geometry} (a) (a different perspective to the oblique diffraction geometry is illustrated in Figure~\ref{fig:oblique_diffraction_geometry} (b)). A collection of possible symmetry-equivalent reflections that can be probed by changing $\omega$ is also extensively discussed in \cite{detlefs2025oblique}.

For an appropriately pre-oriented single crystal, $\mathbf{B}_0$ is constant, and $\mathbf{U}$ is constant for the selected group of symmetry-equivalent reflections for that single crystal. Additionally, in $\mathbf{\Gamma}( \mu, \omega, \chi, \phi)$, $\mu$ is set to zero for the selected reflections to ensure the same illumination volume for different reflections that meet the diffraction condition through $\omega$ rotations. We can choose the orientation of the sample such that $\mu = 0$, and the corresponding $\mathbf{U}$ can be determined accordingly.

Let the selected symmetry-equivalent reflections be denoted as $a_1 = (h_1, k_1, \ell_1)^T$, $a_2 = (h_2, k_2, \ell_2)^T$, and $a_3 = (h_3, k_3, \ell_3)^T$ and their corresponding vectors in the grain coordinate system according to the Eq.~\ref{eq:diffraction_condition_deformed_final_2} are $(h^\prime_1, k^\prime_1, \ell^\prime_1)^T$, $(h^\prime_2, k^\prime_2, \ell^\prime_2)^T$, and $(h^\prime_3, k^\prime_3, \ell^\prime_3)^T$, respectively. Using Eq.~\ref{eq:diffraction_condition_deformed_final_2} for each symmetry-equivalent reflection we can derive a set of 9 equations that comes from Eq.~\ref{eq:diffraction_equation_full_form_single_reflection_h} for each reflection uniquely. This now makes for nine equations to solve the nine unknowns of the $\mathbf{F^{(g)}}$ as shown below.
\begin{equation}
    \resizebox{\textwidth}{!}{$
    \left\{
    \begin{aligned}
        h^\prime_1 T_{11} + k^\prime_1 T_{12} + \ell^\prime_1 T_{13} &= -P^{(1)}_{11}( \omega, \chi, \phi, \lambda) \sin(\theta_1)^2 -P^{(1)}_{12}( \omega, \chi, \phi, \lambda) \cos(\theta_1) \sin(\eta) \sin(\theta_1) +P^{(1)}_{13}( \omega, \chi, \phi, \lambda) \cos(\theta_1) \cos(\eta) \sin(\theta_1) \\
        h^\prime_1 T_{21} + k^\prime_1 T_{22} + \ell^\prime_1 T_{23} &= -P^{(1)}_{21}( \omega, \chi, \phi, \lambda) \sin(\theta_1)^2 -P^{(1)}_{22}( \omega, \chi, \phi, \lambda) \cos(\theta_1) \sin(\eta) \sin(\theta_1) +P^{(1)}_{23}( \omega, \chi, \phi, \lambda) \cos(\theta_1) \cos(\eta) \sin(\theta_1) \\
        h^\prime_1 T_{31} + k^\prime_1 T_{32} + \ell^\prime_1 T_{33} &= -P^{(1)}_{31}( \omega, \chi, \phi, \lambda) \sin(\theta_1)^2 -P^{(1)}_{32}( \omega, \chi, \phi, \lambda) \cos(\theta_1) \sin(\eta) \sin(\theta_1) +P^{(1)}_{33}( \omega, \chi, \phi, \lambda) \cos(\theta_1) \cos(\eta) \sin(\theta_1) \\
        h^\prime_2 T_{11} + k^\prime_2 T_{12} + \ell^\prime_2 T_{13} &= -P^{(2)}_{11}( \omega, \chi, \phi, \lambda) \sin(\theta_2)^2 -P^{(2)}_{12}( \omega, \chi, \phi, \lambda) \cos(\theta_2) \sin(\eta) \sin(\theta_2) +P^{(2)}_{13}( \omega, \chi, \phi, \lambda) \cos(\theta_2) \cos(\eta) \sin(\theta_2) \\
        h^\prime_2 T_{21} + k^\prime_2 T_{22} + \ell^\prime_2 T_{23} &= -P^{(2)}_{21}( \omega, \chi, \phi, \lambda) \sin(\theta_2)^2 -P^{(2)}_{22}( \omega, \chi, \phi, \lambda) \cos(\theta_2) \sin(\eta) \sin(\theta_2) +P^{(2)}_{23}( \omega, \chi, \phi, \lambda) \cos(\theta_2) \cos(\eta) \sin(\theta_2) \\
        h^\prime_2 T_{31} + k^\prime_2 T_{32} + \ell^\prime_2 T_{33} &= -P^{(2)}_{31}( \omega, \chi, \phi, \lambda) \sin(\theta_2)^2 -P^{(2)}_{32}( \omega, \chi, \phi, \lambda) \cos(\theta_2) \sin(\eta) \sin(\theta_2) +P^{(2)}_{33}( \omega, \chi, \phi, \lambda) \cos(\theta_2) \cos(\eta) \sin(\theta_2) \\
        h^\prime_3 T_{11} + k^\prime_3 T_{12} + \ell^\prime_3 T_{13} &= -P^{(3)}_{11}( \omega, \chi, \phi, \lambda) \sin(\theta_3)^2 -P^{(3)}_{12}( \omega, \chi, \phi, \lambda) \cos(\theta_3) \sin(\eta) \sin(\theta_3) +P^{(3)}_{13}( \omega, \chi, \phi, \lambda) \cos(\theta_3) \cos(\eta) \sin(\theta_3) \\
        h^\prime_3 T_{21} + k^\prime_3 T_{22} + \ell^\prime_3 T_{23} &= -P^{(3)}_{21}( \omega, \chi, \phi, \lambda) \sin(\theta_3)^2 -P^{(3)}_{22}( \omega, \chi, \phi, \lambda) \cos(\theta_3) \sin(\eta) \sin(\theta_3) +P^{(3)}_{23}( \omega, \chi, \phi, \lambda) \cos(\theta_3) \cos(\eta) \sin(\theta_3) \\
        h^\prime_3 T_{31} + k^\prime_3 T_{32} + \ell^\prime_3 T_{33} &= -P^{(3)}_{31}( \omega, \chi, \phi, \lambda) \sin(\theta_3)^2 -P^{(3)}_{32}( \omega, \chi, \phi, \lambda) \cos(\theta_3) \sin(\eta) \sin(\theta_3) +P^{(3)}_{33}( \omega, \chi, \phi, \lambda) \cos(\theta_3) \cos(\eta) \sin(\theta_3).
        \end{aligned}
    \right.
    $}
    \label{eq:multiple_reflections}
\end{equation}

In the system of equations in  Eq.~\ref{eq:multiple_reflections}, $P^{(n)}_{ij}$ terms are the components of the $\mathbf{P}^{(n)}$ matrix and these components are defined based on the $\omega, \chi, \phi$ and the $n^{th}$ reflection. The $\mathbf{P}^{(n)}$ matrix has its unique $\omega, \chi, \text{and } \phi$ angles for the reflection $n$ (i.e., $\mathbf{P}^{(n)}$ corresponding to the reflection vector $\vec{a}_n$ and $\phi_n$ and $\chi_n$ are the corresponding $\phi$ and $\chi$ angles for the reflection $\vec{a}_n$). We note, in recent experiments \citep{irvine2025dark},  we have also performed energy scans instead of $\Delta \theta$ scans. In energy scans, $\lambda$ is a function of energy and not a constant. For the purpose of this paper, however, we only consider experiments with $\Delta\theta$  scan where $\lambda$ is taken as a constant. 

For the geometry proposed in \cite{detlefs2025oblique}, the $\eta$ angle remains the same for all the reflections. $\theta_n$ is given by ($\theta_\text{B} + \Delta \theta_n$) for the reflection $\vec{a}_n$, where $\theta_\text{B}$ is the Bragg angle and $\Delta \theta_n$ is the deviation from the Bragg angle. The $\theta_\text{B}$ is the same for all the symmetry-equivalent reflections. By solving the system of non-linear equations in Eq.~\ref{eq:multiple_reflections}, we can reconstruct $\mathbf{T}$ and use the inverse of Eq.~\ref{eq:deformation_gradient_inverse} to calculate $\mathbf{F^{(g)}}$, i.e., $\mathbf{F^{(g)}} = (\mathbf{T}^{T})^{-1}$. Thus, our formalism above enables us the full reconstruction of deformation gradient tensor $\mathbf{F^{(g)}}$. 

\section[Example of Reconstructing the F from DFXM in Aluminum]{Example of Reconstructing the $\mathbf{F^{(g)}}$ from DFXM in Aluminum}
\label{sec:reconstruction_example}

In this section, we demonstrate examples of specific $\mathbf{F^{(g)}}$'s in FCC aluminum
and calculate the angular shifts ($\phi$, $\chi$, and $\Delta \theta$) for all reflections. The $\mathbf{F^{(g)}}$ tensors were selected to represent isolated versions of axial and shear strain, and lattice rotations to reveal how each of them affects the angular shifts. We close the section reconstructing $\mathbf{F^{(g)}}$ using the angular shifts to demonstrate the precision of our approach.


\subsection{Geometry used in the Reconstruction}
\label{Geometry used in the reconstruction}

This example simulates diffraction from FCC aluminum with  the lattice parameter a = 4.0495~$\text{\AA}$  along three $\{2, 0, 2\}$ planes, i.e, $(2, 0, \pm 2)^T$, $(0, \bar{2}, 2)^T$ reflections, rotated about the $(2, \bar{2}, 0)^T$ axis (a.k.a., $\vec{h}_{sym}$). To describe our sample coordinate system using \cite{detlefs2025oblique, poulsen2021geometrical} formalism, we define $\vec{h}_{sym}$ as the \textit{z}-axis, and $\vec{h}_1 = (1, 1, 0)^T$ as the \textit{x}-axis and $\vec{h}_2 = (0, 0, 1)^T$ as the \textit{y}-axis in the sample coordinate system. This allows us to define the $\mathbf{U}$ matrix that transforms from the sample to lab coordinate system as 
\begin{equation}
    \mathbf{U} = \left( \begin{array}{ccc} \frac{1}{\sqrt{2}} & \frac{1}{\sqrt{2}} & 0 \\ 0 & 0 & 1 \\ \frac{1}{\sqrt{2}} & -\frac{1}{\sqrt{2}} & 0 \end{array} \right).
\end{equation}
For full derivation of $\mathbf{U}$ see \ref{appendix1}. As defined in the formalism by \citep{poulsen2021geometrical}, $\mathbf{B}_0$ defines the crystallography of the unit cell as an orthogonal matrix. For aluminum, $\mathbf{B}_0$  is given by,
\begin{equation}
    \mathbf{B}_0 = \frac{2\pi}{a} \left( \begin{array}{ccc} 1 & 0 & 0 \\ 0 & 1 & 0 \\ 0 & 0 & 1 \end{array} \right).
\end{equation}

This work focuses on understanding the crystallography of the deformation and the associated reconstruction.  As such, we do not include the reciprocal-space resolution function that describes the contribution from energy bandwidth and angular divergence to image formation. 

We use the X-ray energy of 19.2 keV for this example. For each reflection, $\theta = 13.03^{\circ}$. The goniometer angle $\omega$  for (202), (20$\bar{2}$), and (0$\bar{2}$2) reflections are 50.35$^{\circ}$, 159.83$^{\circ}$, and -20.17$^{\circ}$, respectively. For each reflection $\eta$ = -59.12$^{\circ}$. By keeping the $\eta$ constant, we do not need to consider the $\Delta \eta$ in the reconstruction and detector position can be kept constant for all the reflections.


Our computations require input parameters in three  categories: X-ray parameters (energy), crystallographic parameters (lattice parameters), and experimental geometry (reflection vectors, $\vec{h}_{sym}$). We construct the $\mathbf{B}$ and $\mathbf{U}$ matrices for solving the system of equations defined in Eq.~\ref{eq:multiple_reflections}. For the forward analysis, we first determine the diffraction angles ($\eta$, $\omega$, and $\theta$) for each reflection for the undeformed crystal. We then apply $\mathbf{F^{(g)}}$ and solve for the required angular shifts ($\phi$, $\chi$, and $\Delta\theta$) to orient the sample and detector for the maximum diffraction intensity in the detector. We accomplish this with the equation system defined in Eq.~\ref{eq:multiple_reflections}. The numerical solution is obtained using the trust region reflective algorithm (SciPy's least squares optimizer with `trf' method), which efficiently handles the bounded nonlinear least squares problem. The initial guess is set to zero for all angular parameters, with bounds defined to allow full $\pm 2 \pi$ radians in each direction. For the reconstruction, we implement the inverse approach, where measured angular shifts are used to recover the unknown $\mathbf{F^{(g)}}$ using a least-squares approach. 

\subsection[Forward Calculating the Angular Shifts from F]{Forward Calculating the Angular Shifts from $\mathbf{F^{(g)}}$}
\label{sec:forward_calculation}

We define six different cases for $\mathbf{F^{(g)}}$'s in the grain coordinate system, corresponding to increasingly complex combinations of axial strain, shear strain and lattice rotations. For each non-zero component of strain or lattice rotation, we select $10^{-3}$ value that describes the limit where the linear approximation $\mathbf{F^{(g)}}=\mathbf{\varepsilon}+\mathbf{w}+\mathbf{I}$ begins to fail.
The six cases are described in Table~\ref{tab:deformation_cases}. 
\begin{sidewaystable}
\begin{table}[H]
    
\centering
\small
\renewcommand{\arraystretch}{1.2}
\begin{tabular}{|>{\centering\arraybackslash}m{2.2cm}
|>{\centering\arraybackslash}m{2.8cm}
|>{\centering\arraybackslash}m{2.8cm}
|>{\centering\arraybackslash}m{2.8cm}
|>{\centering\arraybackslash}m{2.8cm}
|>{\centering\arraybackslash}m{3.2cm}
|>{\centering\arraybackslash}m{3.2cm}|}
\hline
 & \textbf{Case 00}& \textbf{Case 01} & \textbf{Case 02} & \textbf{Case 03} & \textbf{Case 04} & \textbf{Case 05} \\
\hline
$\mathbf{\varepsilon}$ 
& None 
& $E^{(g)}_{xx} = 10^{-3}$ 
& $E^{(g)}_{xy} = 10^{-3}$ 
& \begin{tabular}{@{}c@{}}$E^{(g)}_{xx} = 10^{-3}$\\$E^{(g)}_{xy} = 10^{-3}$\end{tabular}
& None
& \begin{tabular}{@{}c@{}}$E^{(g)}_{xx} = 10^{-3}$\\$E^{(g)}_{xy} = 10^{-3}$\end{tabular} \\
\hline
$\mathbf{w}$ 
& None 
& None 
& None 
& None 
& $R_{x} = 10^{-3}$ rad 
& $R_{x} = 10^{-3}$ rad \\
\hline
\textbf{Description} 
& Undeformed crystal 
& Pure axial strain 
& Pure shear strain 
& Axial and shear strain 
& Pure rotation 
& Axial and shear strain and rotation \\
\hline
$\mathbf{F}^{(g)}_n$ 
& \tiny$\begin{pmatrix} 1.000 & 0.000 & 0.000 \\ 0.000 & 1.000 & 0.000 \\ 0.000 & 0.000 & 1.000 \end{pmatrix}$ 
& \tiny$\begin{pmatrix} 1.001 & 0.000 & 0.000 \\ 0.000 & 1.000 & 0.000 \\ 0.000 & 0.000 & 1.000 \end{pmatrix}$ 
& \tiny$\begin{pmatrix} 1.000 & 0.001 & 0.000 \\ 0.001 & 1.000 & 0.000 \\ 0.000 & 0.000 & 1.000 \end{pmatrix}$ 
& \tiny$\begin{pmatrix} 1.001 & 0.001 & 0.000 \\ 0.001 & 1.000 & 0.000 \\ 0.000 & 0.000 & 1.000 \end{pmatrix}$ 
& \tiny$\begin{pmatrix} 1.000 & 0.000 & 0.000 \\ 0.000 & 1.000 & -0.001 \\ 0.000 & 0.001 & 1.000 \end{pmatrix}$ 
& \tiny$\begin{pmatrix} 1.001 & 0.001 & 0.000 \\ 0.001 & 1.000 & -0.001 \\ 0.000 & 0.001 & 1.000 \end{pmatrix}$ \\
\hline
\end{tabular}
\caption{Deformation gradient tensors ($\boldsymbol{F}_n$) for all six deformation cases. The strain tensor components ($\mathbf{\varepsilon}$) and rotation vector components ($\mathbf{w}$) define the imposed deformation state, progressing from the reference configuration (Case 0) through pure deformations to coupled deformation-rotation states (Case 05). $\boldsymbol{F}_n$ is approximated to the third decimal place.}
\label{tab:deformation_cases}
\end{table}
\end{sidewaystable}

\begin{figure}[H]
    \centering
    \includegraphics[width=\linewidth]{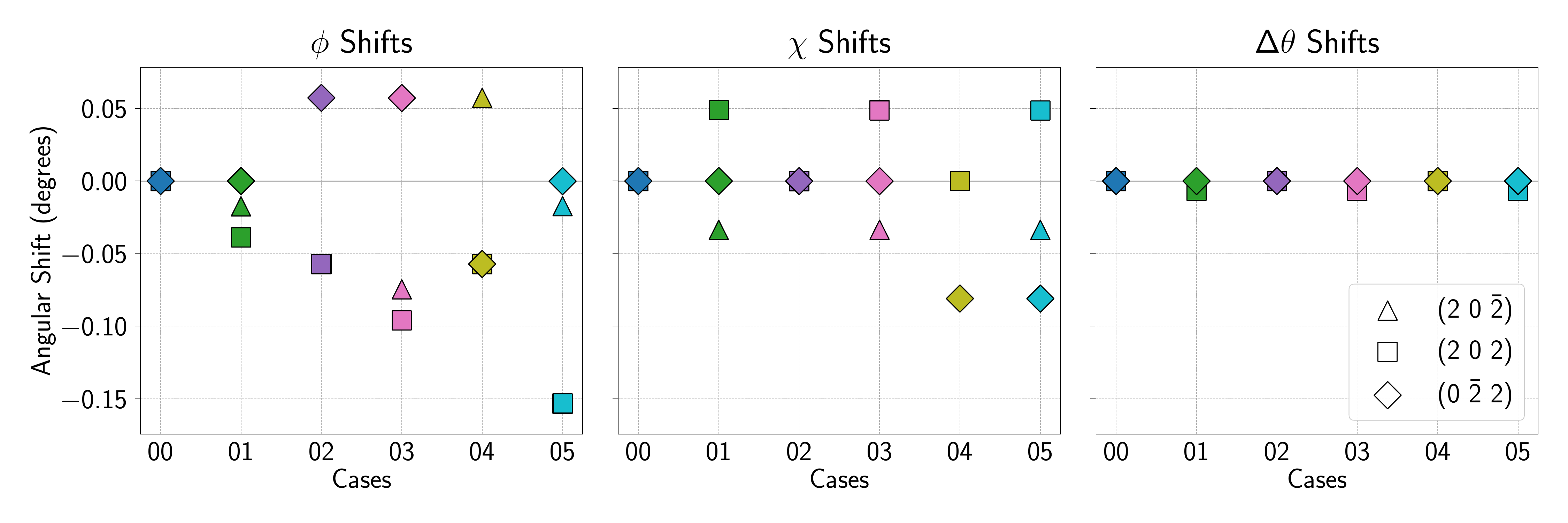}
    \caption{Angular shifts corresponding to the most intense diffraction signal for (2 0 $\bar{2}$), (2 0 2), and (0 $\bar{2}$ 2) reflections for the cases described in Table~\ref{tab:deformation_cases}. Triangles, squares, and rhombuses correspond to (2 0 $\bar{2}$), (2 0 2), and (0 $\bar{2}$ 2) reflections, respectively. The dark blue, green, purple, pink, light green, and light blue markers correspond to Case 00, 01, 02, 03, 04, and 05, respectively.} 
    \label{fig:shifts}
\end{figure}

Using Eq.~\ref{eq:multiple_reflections}, we solve for the angular shifts ($\phi$, $\chi$, and $\Delta\theta$) corresponding to the maximum diffracted intensity for each case in Table~\ref{tab:deformation_cases} by plugging in the values of $\mathbf{F}^{(g)}_n$ and using the least-squares method in Section~\ref{Geometry used in the reconstruction}. The forward-calculated angular shift for each $\mathbf{F}^{(g)}_n$ case is plotted in Figure~\ref{fig:shifts}. 

Figure~\ref{fig:shifts} demonstrates the geometrical relationship between $\mathbf{F^{(g)}}$ and the resulting angular shifts. The angular shifts describe how the reciprocal lattice vector different reflections are modified by each component of $\mathbf{F^{(g)}}$. When the crystal is undeformed ($\mathbf{F^{(g)}}$ = $\mathbf{I}$ as in Case 0), we observe all angular shifts are zero, as shown in the dark blue markers in the Figure~\ref{fig:shifts}.  

Case 01 describes pure axial strain along the \textit{x}-axis ($E^{(g)}_{xx} = 10^{-3}$), resulting in non-zero angular shifts for (2 0 $\bar{2}$) and (2 0 2) reflections. This observation is consistent with the fact that the (2 0 $\bar{2}$) and (2 0 2) ($\vec{Q} = [h, 0, \ell]$) reflections have components along the direction of the strain  (\textit{x}-axis). Both show identical $\Delta\theta = -0.0066^{\circ}$, resulting from an identical change in \textit{d}-spacing due to the axial strain. Notably, these reflections display opposite $\chi$ shifts (-0.034$^{\circ}$ and +0.049$^{\circ}$) as a consequence of their mirror symmetry about the ($h$ 0 0) plane. 
The (0 $\bar{2}$ 2) reflection shows effectively a zero shift because its reciprocal space vector $\vec{Q} = [0, k, \ell]$ is orthogonal to the strain direction (\textit{x}-axis). 

Case 02 describes pure shear strain in the $xy$-plane ($E^{(g)}_{xy} = E^{(g)}_{yx} = 10^{-3}$), resulting in similar magnitudes of $\phi$ ($|\phi| \approx 0.057^{\circ}$) that are negative for (2 0 2) and (2 0 $\bar{2}$) and positive for (0 $\bar{2}$ 2). The signs of $\phi$ arise from the projection of the shear strain along different reflections. $\Delta\theta$ shifts remain negligible (less than ${10^{-5}}^{\circ}$), indicating that shear strain has little effect on the lattice spacing.  

Case 03 is a combination of axial (Case 01) and shear (Case 02) strains along the \textit{x}-axis ($E^{(g)}_{xx} = 10^{-3}$ and $E^{(g)}_{xy} = E^{(g)}_{yx} = 10^{-3}$). The angular shifts observed in Case 03 show a linear superposition of the effects of individual strains from Case 01 and Case 02. For example, $\phi_{\text{Case 03}} \approx \phi_{\text{Case 01}} + \phi_{\text{Case 02}}$ within numerical precision. The additivity in the angular shifts observed in Case 03 is resulting from no coupling between the strains of the order of 10$^{-3}$ in the infinitesimal deformation approximation (Eq.~\ref{eq:deformation_gradient_tensor}). 

Case 04 describes a pure rotation about the \textit{x}-axis ($R^{(g)}_{x} = 10^{-3}$ rad), produces no $\Delta\theta$ shifts. However, the (0 $\bar{2}$ 2) reflection shows a large $\chi$ shift (-0.081$^{\circ}$) compared to the other reflections. This amplification occurs because its reciprocal vector lies entirely in the $yz$-plane, perpendicular to the axis of rotation, resulting in the maximum coupling between the rotation and the diffraction condition. Additionally, there are significant $\phi$ shifts for all the reflections demonstrating the fact that the rotation is sensitive to both $\phi$ and $\chi$ shifts. 
 
Case 05 is a combination of axial and shear strains along the \textit{x}-axis ($E^{(g)}_{xx} = 10^{-3}$ and $E^{(g)}_{xy}$ = $E^{(g)}_{yx} = 10^{-3}$) and pure rotation along the \textit{x}-axis ($R^{(g)}_{x} = 10^{-3}$ rad), the shifts follow the superposition of the shifts from the pure axial and shear strains and the pure rotation. However, the shifts do not follow a linear superposition. The non-linearity demonstrates the onset of finite deformation effects where the multiplicative nature of $\mathbf{F^{(g)}}$ is apparent. The non-linear coupling between the components of strains and rotations  highlights the importance of the full finite deformation formalism for accurately reconstructing $\mathbf{F^{(g)}}$.

These results from angular shifts reveal that each reflection's sensitivity to specific components of $\mathbf{F^{(g)}}$ depends on both its crystallographic orientation and the goniometer configuration. Further, the superposition nature of shifts due to different components of $\mathbf{F^{(g)}}$ is also apparent due to the infinitesimal deformation approximation. We will explore these in more detail in the Section~\ref{Sensitivity analysis}.

\subsection[Reconstructing F from the Angular Shifts]{Reconstructing $\mathbf{F^{(g)}}$ from the Angular Shifts}
\label{reconstructing F from the angular shifts}

Given the goniometer and detector positions corresponding to the maximum intensity, we can reconstruct $\mathbf{F^{(g)}}$ and consequently determine $\mathbf{E}$ and $\mathbf{R}$ using the decomposition. For the 10$^{-3}$ order deformations, one can also use the infinitesimal approximation to calculate $\mathbf{F^{(g)}}$ and use it to approximate the strain and lattice rotation tensors using  Eq. \ref{eq:deformation_gradient_tensor} as discussed in   \ref{sec:comparison_with_infinitesimal_deformation}. 

For the example discussed in this section, we use a precision of four decimal points (i.e., 0.0001 error) for all simulated values that were input for the reconstruction. In this paper, we only present Cases 00 and 05. 

We specifically select Case 00 and Case 05 for our reconstruction in this subsection as the reconstruction in Case 00 represents the ability to recover $\mathbf{F^{(g)}}$ as an identity matrix in the case of the nominal Bragg condition, and Case 05 represents the most complex reconstruction in our example with multiple non-zero components of strain and lattice rotations. 

In Case 00, using the nominal Bragg condition, we are able to reconstruct $\mathbf{F^{(g)}}$ = $\mathbf{I}$. 

In Case 05, we initiate $\mathbf{F^{(g)}} = \mathbf{I}$ for the reconstruction. The initial guesses for each of the component of the $\mathbf{F^{(g)}}$ is updated using the measured angular shifts. The reconstructed $\mathbf{F^{(g)}}$ is then used to calculate the $\mathbf{E}$ and $\mathbf{R}$. Finally, the reconstructed $\mathbf{F^{(g)}}$ matrix are compared with the actual $\mathbf{F^{(g)}}$ matrix used in the simulation. The reconstructed $\mathbf{F^{(g)}}$ and decomposed $\mathbf{E}$ and $\mathbf{R}$ tensors are
\begin{equation}
    \resizebox{\textwidth}{!}{$
    \text{$\mathbf{F_{recon}}$} = \left( \begin{array}{ccc} 1.0009 & 0.0010 & 0.0000 \\ 0.0010 & 0.9999 & -0.0010 \\ 0.0000 & 0.0010 & 0.9998 \end{array} \right), \quad \text{$\mathbf{E_{recon}}$} = \left( \begin{array}{ccc} 0.0009 & 0.0010 & 0.0000 \\ 0.0010 & -0.0001 & 0.0000 \\ 0.0000 & 0.0000 & -0.0002 \end{array} \right), \quad \text{$\mathbf{R_{recon}}$} = \left( \begin{array}{ccc} 1.0000 & 0.0000 & 0.0000 \\ 0.0000 & 1.0000 & -0.0010 \\ 0.0000 & 0.0010 & 1.0000 \end{array} \right).
    $}
\end{equation}

\section{Computing Angular Shifts and Sensitivity Analysis}

The formalism introduced in Section~\ref{Reconstructing the F from DFXM observables} allows a precise calculation of how $\mathbf{F^{(g)}}$ alters the angular shifts required to observe the diffraction, $\phi$, $\chi$, and $\Delta\theta$. This predictive capability enables users to design scan strategies tailored to the expected deformation profiles. In Section~\ref{Deflection analysis}, we illustrate how this formalism facilitates the computation of how $\mathbf{F^{(g)}}$ changes the nominal Bragg condition by the angular shifts for individual reflections. We evaluate the angular response as a function of deformation magnitude by systematically varying strain and lattice rotation, and we explore the trends in these measurable quantities. We build on the deflection analysis in Section~\ref{Sensitivity analysis} by performing a sensitivity analysis to exactly quantify which angular scans ($\phi$, $\chi$, and $\Delta \theta$) are optimal to quantify individual components of $\mathbf{F^{(g)}}$. 

\subsection{Computing the Angular Shifts}
\label{Deflection analysis}

\begin{figure}[H]                                                      
    \centering
     \includegraphics[width=\linewidth]
{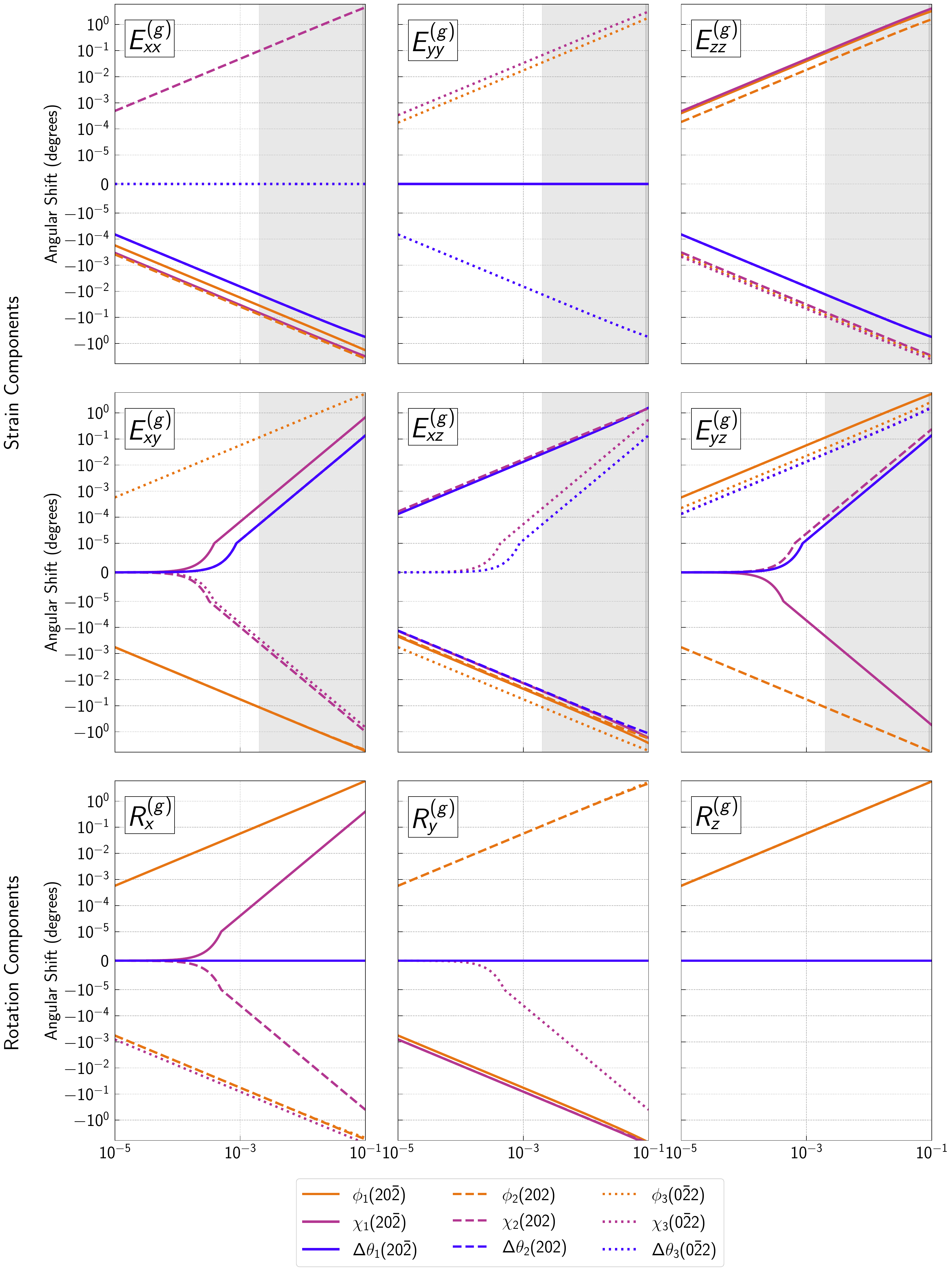}
    \caption{Angular shifts: $\phi$, $\chi$, and $\Delta\theta$ as a function of individual components of strain and rotation for three symmetry-equivalent reflections, $(2, 0, \pm 2)^T$ and $(0, \bar{2}, 2)^T$ in aluminum. The plot uses a symmetric logarithmic (symlog) scale, but converts to linear scaling at a threshold of $10^{-5}$. The white and grey boxes illustrate strain fields characteristic of metals and ceramics, respectively.}
    \label{fig:deflection_anlysis}
\end{figure} 

In Figure~\ref{fig:deflection_anlysis}, we include computed angular shifts required to observe the diffraction from the individual reflections we selected, as discussed in Section~\ref{sec:reconstruction_example}. We start with 10$^{-5}$ strain and increase up to 0.1 strain (i.e., 10\%) and for the rotation, we start from 10$^{-5}$ and increase to 0.1 radians while keeping the other components at their initial positions. This strain range covers the deformation limits of diverse material systems, including the fracture limits of high-strength brittle ceramics (typically $<10^{-3}$ strain), yield on set of ductile metals (e.g., $2 \times 10^{-3}$ strain), the elastic regime of diamond nanostructures under extreme loading conditions (up to $9 \times 10^{-2}$), with the upper bound reaching a maximum reported elastic strain of 13.4\% \citep{nie2019approaching}. We have used a logarithmic scale for different positions the increments and 300 samples for every component of the strain and rotation.

Figure~\ref{fig:deflection_anlysis} shows that each component of the strain and rotation tensors influences angular shifts in a distinct way for the selected reflections. For example, the axial strain $E^{(g)}_{xx}$ affects the angular shifts of the $(202)$ and $(20\bar{2})$ reflections but has no effect on the $(0\bar{2}2)$ reflection, as axial strain along $x_g$-axis ($E^{(g)}_{xx}$) does not project onto the $(0\bar{2}2)$ plane. Whereas, $E^{(g)}_{yy}$ affects only the $(0\bar{2}2)$ reflections, as no other reflection has a projection along $y_g$-direction. Further, $E^{(g)}_{zz}$ affects all three reflections as all three reflections have projection on $z_g$-direction. The off-diagonal strain components ($E^{(g)}_{xy}, E^{(g)}_{xz}, \text{ and }, E^{(g)}_{yz}$) couple more strongly to the orientation angles ($\phi$ and $\chi$) to reorient the shear deformation. The rotations ($R^{(g)}_{x}$, $R^{(g)}_{y}$, and $R^{(g)}_{z}$) exhibit much higher sensitivity than the strain components of the same magnitude, often producing angular shifts that are orders of magnitude larger. This increased sensitivity is due to the direct reorientation of the crystal lattice, which immediately alters the diffraction condition. For the rotation components, $\Delta \theta$ does not show any angular shift at any magnitude since rotation do not affect the \textit{d}-spacing. 

The log-log representation in Figure~\ref{fig:deflection_anlysis} shows two distinct deformation regimes characterized by different power-law dependencies. In the small deformation regime,  the angular shifts exhibit a linear proportionality to the applied deformation. This behavior aligns with the linearized theory of elasticity, where the deformation gradient can be approximated as Eq.~\ref{eq:deformation_gradient_tensor}, showing that angular shifts scale linearly with strain and rotation magnitudes. However, as deformations exceed the small deformation regime, the system transitions into a non-linear regime, where the power-law exponent deviates from unity and becomes deformation-dependent. This transition reflects the breakdown of the small-strain approximation and the emergence of geometric non-linearities. 

The ability to calculate angular shifts is valuable in DFXM, as it shows which reflections and scans are sensitive to specific components of strain or rotation. During an experiment, involving either a single reflection or multiple reflections, it is feasible to calculate the angular shifts of the scans through a range of deformations. The derived angular shifts precisely indicate the movement of the motors needed to probe the shifted diffraction vector.  When targeting a single component of $\mathbf{F^{(g)}}$, users can choose a reflection and scan combination that responds only to that component, minimizing interference from others. Selecting the optimal combination is especially important in time-limited experiments, where selecting the most efficient scan is critical.

The family-dependent nature of these relationships, as evidenced by the different trends observed for the \{400\} family discussed in \ref{Appendix4}, shows the importance of comprehensive pre-experimental modeling to optimize reflection selection for specific deformation studies. This systematic understanding of angular shift behavior across different strain and rotation magnitudes, reflection families, and scan types forms the foundation for quantitative strain and rotation mapping in DFXM experiments.

\subsection{Sensitivity Analysis}
\label{Sensitivity analysis}

In this section we derive how sensitive each reflection is to different components of deformation. To accomplish this, we apply deformations of each types and show an example of how the motors through angular shifts encode that deformation uniquely. We apply a direct and general approach using the finite difference method to assess the absolute sensitivities at specific motor positions. Then we propagate the uncertainty from the accuracy of the motors through linear approximation of Eq.~\ref{eq:multiple_reflections} to quantify the accuracy of the reconstruction. Our approach will help the users to select the optimal reflections to reconstruct the $\mathbf{F^{(g)}}$ with appropriate accuracy for their deformation of interest. 

Usually, the sensitivity of a measured quantity to a parameter of interest is quantified by the dimensionless sensitivity parameter, the relative sensitivity ($\xi$) \citep{chalise2023electron}. In the context of DFXM,  $\xi$ would relate an angular shift (such as $\phi$) to a component of $\mathbf{F^{(g)}}$ (such as $E^{(g)}_{xx}$), with the equation
\begin{equation}
    \xi = \frac{d\phi}{dE^{(g)}_{xx}} \times \frac{E^{(g)}_{xx}}{\phi}.
\end{equation}
The value of $\xi$ is the percent change in $\phi$ that results from a 1\% change in $E^{(g)}_{xx}$. However, $\xi$ is not suitable for sensitivity analysis in DFXM because the angular shifts at the nominal Bragg condition are 0, making the normalization term $\frac{E^{(g)}_{xx}}{\phi}$ undefined.

For the case of DFXM, we define an absolute sensitivity ($\nu$) using the forward finite-difference approximation \citep{lenhart2002comparison}. The forward finite-difference approximation quantifies how sensitive each angular shift is to the different components of $\mathbf{F^{(g)}}$. Using $\phi$ and $E^{(g)}_{xx}$ as examples, the $\nu_{\phi,E^{(g)}_{xx}}$ describes the change in $\phi$ per unit change in $E^{(g)}_{xx}$, assuming a linear relationship between $\phi$ and increments in $E^{(g)}_{xx}$. This relationship holds true for small deformations, and is calculated as
\begin{equation}
        \nu_{\phi,E^{(g)}_{xx}} = \frac{\Delta \phi}{\Delta E^{(g)}_{xx}},
\end{equation}
where $\Delta E^{(g)}_{xx}$ is the strain increment considered and $\Delta \phi$ is the resulting change in $\phi$. 

\begin{figure}[H]
    \centering
    \includegraphics[width=\linewidth]{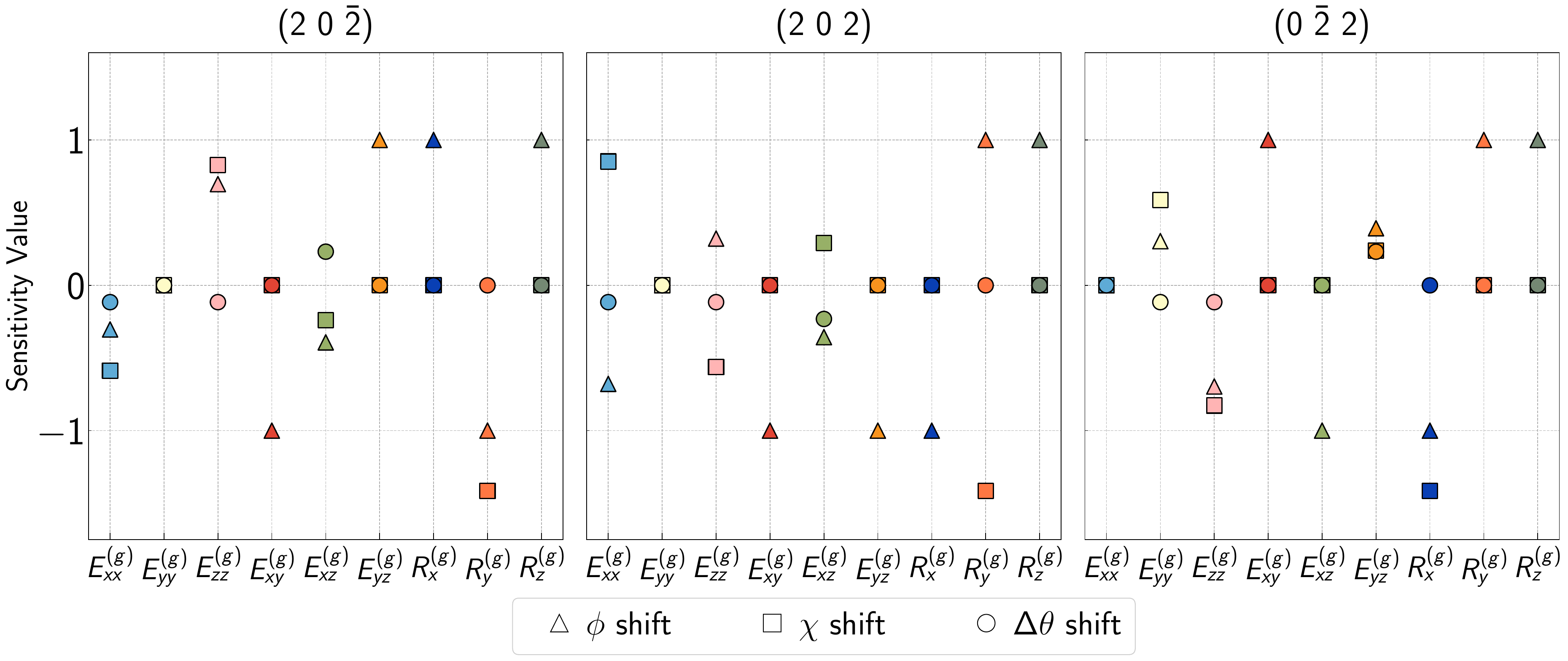}
    \caption{Absolute sensitivity analysis: Absolute sensitivity of angular shifts ($\phi$, $\chi$, and $\Delta \theta$) for selected symmetry-equivalent reflections (202), (20$\bar{2}$), and (0$\bar{2}$2), calculated with respect to infinitesimal deformation of 10$^{-5}$ from the undeformed state ($\mathbf{F^{(g)}} = \mathbf{I}$). The sensitivity values corresponding to the strain components are expressed in units of radians, whereas those associated with the rotation components are dimensionless. Triangles, squares, and circles correspond to $\phi$, $\chi$, and $\Delta \theta$ shifts, respectively. The different colors correspond to the different components of the strain and rotation tensors.}
    \label{fig:absolute_sensitivity}
\end{figure}

We demonstrate the absolute sensitivity of different motor positions to the components of strain and rotation for the three symmetry-equivalent reflections of discussed above $\{202\}$ family in Figure~\ref{fig:absolute_sensitivity}. These absolute sensitivities are calculated by setting each independent components of the strain and rotation tensors to $10^{-5}$ with all other components held at zero. For example, to compute $\nu_{\phi_1,E^{(g)}_{xx}}$, we apply a small perturbation $\Delta E^{(g)}_{xx}$ = $10^{-5}$ to the strain component while keeping all other strain and rotation components at zero. We then calculate the resulting change in $\phi_1$, i.e., $\phi$ for the (202) reflection, using Eq.~\ref{eq:multiple_reflections} in our forward model. From the calculation, we obtain $\nu_{\phi_1,E^{(g)}_{xx}}$ = -0.68, indicating $\phi_1$ decreases by 0.68 degrees over an increment of 10$^{-5}$ in $E^{(g)}_{xx}$. 

From Figure~\ref{fig:absolute_sensitivity}, we observe  $\phi$ and $\chi$ motors show significantly higher sensitivities ($|\nu| \geq 0.3$) compared to $\Delta \theta$, which remains consistently below 0.2 across all strains and rotations. The low sensitivity of $\Delta \theta$ arises because $\Delta \theta$ represents the change in the Bragg angle, which is primarily sensitive to the changes in the lattice spacing (\textit{d}-spacing variations). For a small magnitude of strain, the relationship $\Delta \theta / \theta \approx - \Delta d / d$ implies that $\Delta \theta$ changes are directly proportional to the change in the lattice parameter. If the magnitude of the strain is small, the change in the lattice-spacing and thereby the change in the $\Delta \theta$ is small. In the Ewald's sphere formalism $\phi$ and $\chi$ motors rotate the sphere, whereas the $\Delta \theta$ motor changes the length of the diffraction vector, i.e., changing the diffraction condition to a sphere of different radius (see Figure 8 in \cite{poulsen2017x}).

For the diagonal components of $\mathbf{F^{(g)}}$ (i.e., $E^{(g)}_{xx}, E^{(g)}_{yy}, \text{ and } E^{(g)}_{zz}$), the sensitivity is non-zero only if the \textit{i}-th component of the reciprocal lattice vector is non-zero. For example, $E^{(g)}_{xx}$ affects the angular shifts of the $(202)$ and $(20\bar{2})$ reflections, but not the $(0\bar{2}2)$ reflection.

For an infinitesimal deformation (Eq.~\ref{eq:deformation_gradient_tensor}), the expected angular shift for a given reflection is a linear function of the strain and rotation components of $\mathbf{F^{(g)}}$, represented by $\varrho$, and its absolute sensitivity. This relationship takes the form
\begin{equation}
    \Delta \psi_i \approx \sum_{j=1}^{9} \nu_{\psi_i,\varrho_j} \Delta \varrho_j = \vec{\nu_i} \cdot \Delta \vec{\varrho},
    \label{eq:absolute_sensitivity_linear}
\end{equation}
where $\Delta\psi_i$ is the change in the angular shift for the $i$-th reflection and $\Delta \varrho_j$ represents the change in the $j$-th component of the strain and rotation tensors. The absolute sensitivity matrix ($\boldsymbol{\nu}$) is a 9 $\times$ 9 matrix where element $\boldsymbol{\nu}_{ij}$ quantifies the absolute sensitivity of the $i$-th angular shift to the $j$-th strain and rotation component of the $\mathbf{F^{(g)}}$. Rows correspond to the 9 angular shifts (3 shifts per reflection for 3 reflections), while columns correspond to the 9 strain and rotation component.



During experimental planning, users can construct $\boldsymbol{\nu}$ for a candidate set of reflections to explicitly predict the angular shift range. This is an indication of which reflection is more sensitive for a deformation of interest. When a linear relationship exists between the angular shifts and the components of $\mathbf{F^{(g)}}$ (Eq.~\ref{eq:absolute_sensitivity_linear}), the sensitivity values described in $\boldsymbol{\nu}$ can be used to calculation of strain and rotation components. The above mentioned calculation can allow us a way of simply reconstructing the $\mathbf{F^{(g)}}$ using a linear set of transformations in contrast to the more complex, but accurate form shown in Eq.~\ref{eq:multiple_reflections}. This relationship is given by
\begin{equation}
    \Delta \vec{\varrho} \approx \boldsymbol{\nu}^{-1} \cdot \Delta \vec{\psi},
    \label{eq:inverse_sensitivity_linear}
\end{equation}
where $\Delta \vec{\psi}$ is the vector of angular shifts ($\phi_1, \chi_1, \Delta \theta_1, \phi_2, \chi_2, \Delta \theta_2, \ldots$), $\Delta \vec{\varrho}$ is the vector of strain and rotation components, and $\boldsymbol{\nu}^{-1}$ is the Moore-Penrose pseudoinverse\footnote{Small measurement errors get amplified enormously through regular matrix inversion, leading to unstable strain reconstructions. The pseudoinverse provides better numerical stability.} of the absolute sensitivity matrix\footnote{The computation for determining the absolute sensitivity matrix is detailed in Section 5.2 of the accompanying Python implementation.} relating the angular shifts to the strain and rotation components.

For Case 05 ($E^{(g)}_{xx} = E^{(g)}_{xy} = 0.001$ and $R^{(g)}_{x} = 0.001$ rad) introduced in Section~\ref{sec:forward_calculation}, we consider experimental angular shifts measured with resolution $\phi_{\text{Res}} = \chi_{\text{Res}} = 10^{-3}$ degrees and $\Delta \theta_{\text{Res}} = 10^{-4}$ degrees that have been demonstrated at beamline ID03 \citep{isern2024esrf}. Using Eq.\ref{eq:inverse_sensitivity_linear} and $\boldsymbol{\nu}$ constructed with infinitesimal perturbations of $10^{-5}$ (Figure~\ref{fig:absolute_sensitivity}, numerical values in \ref{absolute_sensitivity_202}), we can construct the $\Delta \vec{\varrho}$. The calculated $\Delta \vec{\varrho}$ for Case 05 is given by
\begin{equation}
    \Delta \vec{\varrho}
    \approx
    \begin{pmatrix}
        E^{(g)}_{xx} \approx 0.001\\
        E^{(g)}_{yy} \approx 0.000\\
        E^{(g)}_{zz} \approx 0.000\\
        E^{(g)}_{xy} \approx 0.001\\
        E^{(g)}_{xz} \approx 0.000\\
        E^{(g)}_{yz} \approx 0.000\\
        R^{(g)}_{x} \approx  0.001\\
        R^{(g)}_{y} \approx 0.000\\
        R^{(g)}_{z} \approx 0.000
    \end{pmatrix} \approx  
    \boldsymbol{\nu}^{-1}
    \cdot
    \begin{pmatrix}
        -0.017 \\
        -0.034 \\
        -0.0066 \\
        -0.153 \\
        0.049 \\
        -0.0066 \\
        -0.000 \\
        -0.081 \\
        0.0000
    \end{pmatrix}.
\end{equation}

As demonstrated above for Case 5, users can use this approach described in Eq.~\ref{eq:inverse_sensitivity_linear}, to be able to compute the angular shifts for any arbitrary strain and rotation tensor fields. In a DFXM imaging experiment, we would recommend that users take this point-based approach and plug the limiting values to understand how different deformation fields that are bounded within the set that they might expect would manifest in the samples. For example, if users are studying metals, they would expect a strain fields with values in the order of $10^{-3}$, as we plotted in Figure~\ref{fig:deflection_anlysis}. 

\section{Uncertainty Quantification}

In this section, we extend our sensitivity analysis to estimate the uncertainty in the reconstructed $\mathbf{F^{(g)}}$ (through $\Delta \vec{\varrho}$) for the given reflections. We anticipate that this section will help users to select the appropriate set of symmetry-equivalent reflections for their desired precision of $\mathbf{F^{(g)}}$ reconstruction.  

From Eq.~\ref{eq:absolute_sensitivity_linear}, we can quantify how uncertainty in the measured angular shifts propagate to uncertainty in the values of the reconstructed $\mathbf{F^{(g)}}$. Then, the Eq.~\ref{eq:absolute_sensitivity_linear} is modified as
\begin{align}
    (\Delta \vec{\psi} \pm \delta\vec{\psi}) &\approx \nu \cdot (\Delta \vec{\varrho} \pm \delta\vec{\varrho}), \\
    \Delta \vec{\psi^\prime} &\approx \nu \cdot \Delta \vec{\varrho^\prime},
    \label{eq:error_in_angular_shifts}
\end{align}
where $\Delta \vec{\psi^\prime} = \Delta \vec{\psi} \pm \delta\vec{\psi}$ is the modified angular shift vector due to the uncertainty in the angular shift measurements and $\Delta \vec{\varrho^\prime} = \Delta \vec{\varrho} \pm \delta\vec{\varrho}$ is the modified strain and rotation components due to the uncertainty in the measured angular shift. Here, $\delta\vec{{\psi}}$ and $\delta\vec{{\varrho}}$ are the uncertainty vectors for  $\Delta \vec{\psi}$ and $\Delta \vec{\varrho}$, respectively. 

Therefore, the uncertainty in the reconstruction of the all the components of the $\Delta \vec{\varrho}$ can be calculated using standard error propagation\footnote{The computation for determining the absolute sensitivity matrix is detailed in Section 6 of the accompanying Python implementation.}. This is given by
\begin{equation}
    {\delta}\vec{{\varrho}} \approx \boldsymbol{\nu}^{-1} \cdot \delta\vec{\psi}.
\end{equation}
For example, in Case 05 described in Section~\ref{sec:forward_calculation}, the uncertainty in the reconstruction of the $\Delta \vec{\varrho}$ components when the angular shifts are measured with the resolution of $\phi_{\text{Res}} = \chi_{\text{Res}} = 10^{-3}$ degrees and $\Delta \theta_{\text{Res}} = 10^{-4}$ degrees can be given by
\begin{equation}
    \begin{pmatrix}
        \delta{E^{(g)}_{xx}} \\
        \delta{E^{(g)}_{yy}} \\
        \delta{E^{(g)}_{zz}} \\
        \delta{E^{(g)}_{xy}} \\
        \delta{E^{(g)}_{xz}} \\
        \delta{E^{(g)}_{yz}} \\
        \delta{R^{(g)}_{x}} \\
        \delta{R^{(g)}_{y}} \\
        \delta{R^{(g)}_{z}} 
    \end{pmatrix} =  \boldsymbol{\nu}^{-1} 
    \begin{pmatrix} 
    \delta{\phi_{(2 0 \bar{2})}} \\ 
    \delta{\chi_{(2 0 \bar{2})}} \\ 
    \delta{\Delta \theta_{(2 0 \bar{2})}} \\ 
    \delta{\phi_{(2 0 2)}} \\ 
    \delta{\chi_{(2 0 2)}} \\ 
    \delta{\Delta \theta_{(2 0 2)}} \\ 
    \delta{\phi_{(0 \bar{2} 2)}} \\ 
    \delta{\chi_{(0 \bar{2} 2)}} \\ 
    \delta{\Delta \theta_{(0 \bar{2} 2)}} 
    \end{pmatrix} 
    =   \begin{pmatrix} 
        \pm 7.41 \times 10^{-6} \\
        \pm 7.41 \times 10^{-6} \\
        \pm 7.67 \times 10^{-6} \\
        \pm 2.69 \times 10^{-7} \\
        \pm 4.12 \times 10^{-12} \\
        \pm 1.06 \times 10^{-11} \\
        \pm 1.09 \times 10^{-5} \\
        \pm 1.38 \times 10^{-5} \\
        \pm 1.75 \times 10^{-5}
    \end{pmatrix}.
    \label{error_for_case5}
    \end{equation}
Here, the $\delta{\phi} = \delta{\chi} = 10^{-3}$ degrees and $\delta{\Delta \theta} = 10^{-4}$ degrees.
From Eq.~\ref{eq:absolute_sensitivity_linear} and Eq.~\ref{eq:error_in_angular_shifts}, we can calculate the total uncertainty  in the reconstruction of the $\Delta \vec{\varrho}$. The total uncertainty is given by (refer to \ref{total_uncertainity})
\begin{equation}
    \frac{\left\lVert{\delta}\vec{{\varrho}}\right\rVert}{\left\lVert\Delta \vec{\varrho}\right\rVert} \leq \left\lVert \boldsymbol{\nu} \right\rVert \left\lVert \boldsymbol{\nu}^{-1} \right\rVert \frac{\left\lVert \delta \vec{\psi} \right\rVert}{\left\lVert \Delta \vec{\psi} \right\rVert}
\end{equation}
where $\left\lVert \ldots \right\rVert$ is the magnitude of the matrix (We used 2-norm for the calculation). $\left\lVert \boldsymbol{\nu} \right\rVert \left\lVert \boldsymbol{\nu}^{-1} \right\rVert$ is the condition number of the sensitivity matrix ($\kappa(\boldsymbol{\nu})$). The theoretical upper bound to the error in the reconstruction is given by $\kappa(\boldsymbol{\nu}) \frac{\left\lVert \delta \vec{\psi} \right\rVert}{\left\lVert \Delta \vec{\psi} \right\rVert}$.

For the example shown in Case 05, the theoretical upper bound on the relative uncertainty for the reconstruction is 25.83\% and the actual relative error calculated using the error in Eq.~\ref{error_for_case5} is 0.05\%. In comparison, for \{113\} family of symmetry-equivalent planes, the relative uncertainty for the reconstruction is 18.57\%  and for \{400\} it is 10.92\%. This suggests that the \{400\} planes are the the best-suited family of planes for the reconstruction of $\mathbf{F^{(g)}}$.

The uncertainty values in Eq.~\ref{error_for_case5} demonstrate that the asymmetric uncertainty for the motor shifts propagates through Eq.~\ref{eq:error_in_angular_shifts} to create uncertainties in $\mathbf{F^{(g)}}$ that differ by orders of magnitude. The choice of symmetry-equivalent reflections changes which motor position uncertainties cause the highest uncertainty in $\mathbf{F^{(g)}}$. We use the parameter $\kappa(\boldsymbol{\nu})$ to quantify the spread of uncertainty values across all components of $\mathbf{F^{(g)}}$, i.e., through the components of $\vec{{\delta}{\varrho}}$, due to the uncertainty in measured angular shifts, $\vec{{\delta}{\psi}}$. If $\kappa(\boldsymbol{\nu}) = 1$ all uncertainties are equal in $\delta\vec{\varrho}$; this is the minimum possible value of $\kappa(\boldsymbol{\nu})$. For values of $\kappa(\boldsymbol{\nu})>>1$, the spread of uncertainty is large. High $\kappa(\boldsymbol{\nu})$ for a selected set of symmetry-equivalent reflections indicates that small measurement errors from the angular motors, $\chi,\phi, \text{ and } \Delta \theta$, will result in large errors in the reconstructed $\mathbf{F^{(g)}}$ (based on the values of $\nu$).

For the selected reflections (202, 20$\bar{2}$, and 0$\bar{2}$2) with infinitesimal deformation of 10$^{-5}$ from the undeformed state ($\mathbf{F^{(g)}} = \mathbf{I}$), $\kappa(\boldsymbol{\nu})$ = 20.90. For reference, the symmetry-equivalent reflection families \{113\} and \{400\} for the same sample described in Section~\ref{Geometry used in the reconstruction}, the $\kappa(\boldsymbol{\nu})$ values are 11.13 and 7.69, respectively. These values corroborate our findings in Section~\ref{Deflection analysis} that \{400\} planes give the most precise reconstruction of $\mathbf{F^{(g)}}$. The \{400\} planes ultimately have higher precision in $\mathbf{F^{(g)}}$ reconstruction because they are higher-order reflections that exhibit an increased sensitivity of $\Delta \theta$ to the components of the $\mathbf{F^{(g)}}$. Additionally, zero values for most of the Miller indices in the \{400\} family of planes reduces the coupling between the components of $\mathbf{F^{(g)}}$ from the $\phi$ and $\chi$ shifts, reducing the propagation of uncertainty. 

Beyond the linear approximation approach in sensitivity analysis shown in this work, we note that the linearity breaks down for large magnitudes of deformation. While these values are appropriate for metals, we note the need for a more careful analysis of the interaction effects of deformation components in higher-strain studies (e.g., ceramics, shock, etc.). We include a global variance-based Sobol sensitivity analysis in \ref{Appendix5} to address the interaction effects.

\section{Discussion}

%
 

The formalism we present in this work demonstrates a step towards using DFXM to map the full $\mathbf{F}(x,y,z)$ field that is important across many fields \citep{callister2022fundamentals}. In crystal plasticity, the deformation of materials can only be expressed in full (i.e., including symmetric and antisymmetric deformation components) using $\mathbf{F}(x,y,z)$. While atomistic models are able to define how single defects describe those deformations, the $\mathbf{F^{(g)}}$ field is required to up-scale those defect populations to bulk deformations. Plastic deformation limits the magnitude of strain fields, however, limited ability for deformation in ceramics causes those studies to include significantly higher strain fields that drive fracture, fatigue, etc. 
As such, our formalism provides a means to quantitatively upscale dislocation dynamics to larger scales using DFXM measurements, and to measure strain fields in important applications beyond metals, including semiconductor technologies \citep{tanner2021x}, energy storage technology \citep{yildirim2024understanding}, and shock deformations \citep{katagiri2023transonic}. 


Our formalism for the full reconstruction of the $\mathbf{F^{(g)}}$ builds on \cite{detlefs2025oblique}, who develop a method to measure three symmetry-equivalent reflections while maintaining a constant illumination volume. While their work focuses on achieving this geometry, our contribution extends the mathematical framework to enable full tensor reconstruction. Additionally, the expression in Eq.~\ref{eq:diffraction_equation_full_form_single_reflection_h} allows the calculation of angular shift and sensitivity from a single reflection, without requiring measurements from all three reflections. The recent work by \cite{henningsson2025towards} further who proposes a regression-based method to fit the $\mathbf{F^{(g)}}$ to DFXM data from three reflections. For full tensor reconstruction, their method should yield equivalent results to ours. Our analytical approach simplifies the computation to allow for full sensitivity analysis, which is essential for optimizing experimental design and quantification of errors.

An important finding in this work is the sensitivity analysis and uncertainty estimation, which can guide planning for experiments (multi- and single-reflection) and interpretation of data for DFXM studies. 
Researchers seeking to select the most effective reflections and motor scans to study their specific deformation fields of interest may use our methods in Section~\ref{Sensitivity analysis} to quantitatively link the measurable angular shifts to the underlying tensor components. 
Our uncertainty propagation then predicts the achievable precision for each reconstructed component of $\mathbf{F^{(g)}}$ -- propagating how motor accuracy, scan resolution, and choice of symmetry-equivalent reflections influence the precision of strain or rotation fields. 
Together, these analyses allow researchers to tailor their experimental geometry and measurement strategy for both the optimal information content and minimal uncertainty along the parameters of interest. This further enables users to prioritize different tradeoffs during limited synchrotron time, e.g., higher precision in specific components of strain/rotation at the cost of others, vs higher precision in all motors but at a lower value than is overall possible. Our work facilitates more reliable comparison between experiment and simulation. 

In this study, we do not consider the reciprocal-space resolution function resulting from the bandwidth and focusing of the incident X-ray beam, or the acceptance angle of the objective lens. For a well-defined reciprocal-space resolution that can be approximated by a Gaussian distribution, our approach is still appropriate so long as the sampling in reciprocal space is sufficient. For experiments in which the reciprocal-space sampling is insufficient, however, the reconstructed values may be less accurate than we predict in Section 5.3. We envision the current formalism can be extended to include the reciprocal-space resolution function by considering the convolution of the diffracted intensity with the resolution function. Integrating the effect of the resolution function is beyond the scope of the current work, but may be included in future publications.

\section{Conclusions}

We have developed a theoretical and computational framework that enables quantitative prediction of DFXM sensitivity to bulk-crystal mechanics both for data analysis and experimental planning/predictions. We analytically derive and validate the complete reconstruction of the $\mathbf{F^{(g)}}$, incorporating finite-deformation effects that can directly resolve all strain and lattice-rotation components at the voxel level. We extend DFXM theory beyond infinitesimal strain limitations to accurately analyze extreme-loading scenarios and/or highly strained materials and use this formalism to devise a forward model predicting angular shifts for a prescribed $\mathbf{F^{(g)}}$ distribution. We use this formalism to present a sensitivity analysis that can quantify the information content for a given set of reflections and scan axes, allowing us to quantify which specific deformation fields are most relevant to experimental observables. Finally, we implement closed-form uncertainty propagation to quantitatively relate the goniometer accuracy/precision to the confidence intervals for reconstructed tensor elements, offering guiding principals for both experimental design and stage selection for beamline construction (e.g., at XFELs). These contributions collectively enable high-precision DFXM for mapping three-dimensional strain and rotation fields during dynamic processes, with relevance across metals, ceramics, and ultrafast processes. 



\section*{Acknowledgments}
LDM and KB were supported by the Army Research Laboratory and the Army Research Office under contract/grant No. W911NF-24-2-0085.
D.C. was supported by the Department of Energy, Office of Science, Basic Energy Sciences, Materials Sciences and Engineering Division, under Contract DEAC02-76SF00515. 

\newpage

\begin{myappendix}

\section[F for Finite Deformations]{$\mathbf{F^{(g)}}$ for Finite Deformations}
\label{appendix-finitedeformation}

In this Appendix, we derive the rotation and finite strain tensors from the $\mathbf{F^{(g)}}$ and compare the finite deformation results with the infinitesimal approximation. 

\subsection[Decomposition of the F]{Decomposition of the $\mathbf{F^{(g)}}$}

In this appendix, we display the mathematical derivation of the rotation tensor from the $\mathbf{F}^{(g)}$ tensor described in the main paper. The $\mathbf{F^{(g)}}$ decomposes multiplicatively into a rotation tensor $\mathbf{R}$ and two stretch tensors:
\begin{equation}
    \mathbf{F^{(g)}} = \mathbf{R_1} \mathbf{V_1} = \mathbf{V_2} \mathbf{R_2},
\end{equation}
where we define 
\begin{itemize}[noitemsep,topsep=0pt]
    \item $\mathbf{R}$: Orthogonal rotation tensor ($\mathbf{R}^\mathsf{T} \mathbf{R} = \mathbf{I}$),
    \item $\mathbf{V_1}$: Right stretch tensor (symmetric, positive definite), Lagrangian (material) description of stretch,
    \item $\mathbf{V_2}$: Left stretch tensor (symmetric, positive definite), Eulerian (spatial) description of stretch.
\end{itemize}

The right Cauchy-Green deformation tensor $\mathbf{C}$ is defined as
\begin{equation}
    \mathbf{C} = \mathbf{F^{(g)}}^\mathsf{T} \mathbf{F^{(g)}} = \mathbf{U}^2.
\end{equation}
Similarly, the values of $\mathbf{U}$ are then computed via singular-value decomposition (SVD) of $\mathbf{C}$, using
\begin{align}
    \mathbf{C} &= \mathbf{Q} \boldsymbol{\Lambda} \mathbf{Q}^\mathsf{T} \quad \text{(diagonalization)}, \\
    \mathbf{U} &= \mathbf{Q} \sqrt{\boldsymbol{\Lambda}} \mathbf{Q}^\mathsf{T},
\end{align}
where $\boldsymbol{\Lambda}$ contains eigenvalues of $\mathbf{C}$ (squares of principal stretches $\lambda_i^2$), and $\mathbf{Q}$ holds eigenvectors. The rotation tensor is then given by
\begin{equation}
    \mathbf{R} = \mathbf{F^{(g)}} \mathbf{U}^{-1}.
\end{equation}

\subsection{Finite Strain Tensor}

We now derive the principal stretches from the Green-Lagrange strain tensor which is used in Section~\ref{reconstructing F from the angular shifts}. The Green-Lagrange strain tensor $\mathbf{E}$ (Lagrangian strain measure) is
\begin{equation}
    \mathbf{E} = \frac{1}{2} (\mathbf{C} - \mathbf{I}) = \frac{1}{2} (\mathbf{U}^2 - \mathbf{I}),
\end{equation}
with principal strains
\begin{equation}
    E_i = \frac{1}{2} (\lambda_i^2 - 1) \quad (i=1,2,3),
\end{equation}
where $\lambda_i$ are eigenvalues of $\mathbf{U}$ (principal stretches) \citep{bonet1997nonlinear}.

\subsection{Finite Lattice Rotation Tensor}

The rotation tensor $\mathbf{R}$ from decomposition describes the rigid-body rotation of the material lattice. The finite lattice rotation tensor $\boldsymbol{\omega}$ is obtained via the matrix logarithm
\begin{equation}
    \mathbf{w} = \log(\mathbf{R}),
\end{equation}
where $\mathbf{w}$ is skew-symmetric ($\mathbf{w}^\mathsf{T} = -\mathbf{w}$). This maps $\mathbf{R}$ (Lie group $\text{SO}(3)$) to its Lie algebra $\mathfrak{so}(3)$, with $\boldsymbol{\omega}$ encoding the rotation axis $(w_1,w_2,w_3)$ and angle $\theta = \|\mathbf{w}\|$ \citep{cardoso2010exponentials}.

\subsection{Comparison with infinitesimal deformation}
\label{sec:comparison_with_infinitesimal_deformation}

Let us consider Case 05 from  Section~\ref{sec:reconstruction_example} and compare the results with the infinitesimal deformation. Conversion of $\mathbf{R}$ to $\mathbf{w}$ is done using the matrix logarithm, and the skew-symmetric tensor is given by:
\begin{equation}
    \text{$\mathbf{w}$} = \left( \begin{array}{ccc} 0.0000 & 0.0000 & 0.0000 \\ 0.0000 & 0.0000 & -0.0010 \\ 0.0000 & 0.0010 & 0.0000 \end{array} \right)
\end{equation}

Now, using the infinitesimal deformation (Eq.~\ref{eq:deformation_gradient_tensor}) and the symmetric strain tensor (Eq.~\ref{eq:strain_limit}), we can compute the infinitesimal $\mathbf{F^{(g)}}$, which in this case, is equal to the reconstructed $\mathbf{F^{(g)}}$ in Section~\ref{sec:reconstruction_example}:
\begin{equation}
    \text{$\mathbf{F_{inf}}$} = \mathbf{I} + \mathbf{w} + \mathbf{\varepsilon} = \left( \begin{array}{ccc} 1.0009 & 0.0010 & 0.0000 \\ 0.0010 & 0.9999 & -0.0010 \\ 0.0000 & 0.0010 & 0.9998 \end{array} \right) = \mathbf{F_{recon}}.
\end{equation}

\section{Orientation matrix}

\subsection{Determining the U matrix for simulations}
\label{appendix1}

The orthogonal matrix $\mathbf{U}$ describes how the crystal is mounted on the innermost goniometer axis (refer to Eq.~\ref{eq:diffraction_vector_sample}). Typically, the $\mathbf{UB}$ matrix is calculated by the diffractometer software. For the purpose of simulations, however, the $\mathbf{U}$ matrix for a given $\mathbf{B}$ matrix and symmetry axis can be determined as follows:

\begin{enumerate}
\item As the $\mathbf{U}$ matrix remains to be determined, we perform most calculations in the \emph{grain coordinate system}.

\item Choose the Miller indices of the symmetry direction,
\begin{align}
    \vec{h}_\sym
    &=
    \begin{pmatrix}
        H_\sym \\ K_\sym \\ L_\sym
        \end{pmatrix}.
\end{align}

Let $\vec{g}_{\sym,\grain} = \mathbf{B} \cdot \vec{h}_\sym$ such that $\vec{g}_{\sym,\diff} = \mathbf{U} \cdot \vec{g}_{\sym,\grain}$. As $\mathbf{U}$ is orthogonal, $\left|\vec{g}_{\sym,\grain}\right| = \left|\vec{g}_{\sym,\diff}\right|$.
We want the symmetry axis to be along the $z$-axis, $\vec{g}_{\sym,\diff} = \left| \vec{g}_\sym \right| \hat{z}_\lab$.

\item Choose two other directions $\vec{h}_1$ and $\vec{h}_2$ such that
$\vec{g}_{\sym,\grain} \perp \vec{g}_{1,\grain}$ and $\vec{g}_{2,\grain} = \vec{g}_{\sym,\grain} \times \vec{g}_{1,\grain}$, where $\vec{g}_{1,\grain} = \mathbf{B} \cdot \vec{h}_1$ and $\vec{g}_{2,\grain} = \mathbf{B} \cdot \vec{h}_2$.

For example, $\vec{h}_1$ can be given by a facet of the sample, known direction from prior characterization, etc.
In the absence of prior information, chose an arbitrary direction perpendicular to $\vec{h}_\sym$.

We want $\mathbf{U} \cdot \vec{g}_{1,\grain}$ to be along the $\hat{x}$-axis, and $\mathbf{U} \cdot \vec{g}_{2,\grain}$ along the $\hat{y}$-axis, $\mathbf{U}\cdot \vec{g}_{1,\grain} = \left| \vec{g}_{1,\grain} \right| \hat{x}_\lab$ and
$\mathbf{U} \cdot \vec{g}_{2,\grain} = \left| \vec{g}_{2,\grain} \right| \hat{y}_\lab$.

\item Group these three vectors into a matrix.
Let
\begin{align}
    \mathbf{H}_\sym
    &=
    \begin{pmatrix} \vec{h}_1, \vec{h}_2, \vec{h}_\sym \end{pmatrix}
    \\
    \mathbf{N}
    &=
    \begin{pmatrix}
    \left|\vec{g}_{1,\grain} \right| \hat{x}_\lab,
    \left|\vec{g}_{2,\grain} \right| \hat{y}_\lab,
    \left|\vec{g}_{\sym,\grain} \right| \hat{z}_\lab
    \end{pmatrix}
    \\
    &=
    \begin{pmatrix}
        \left|\vec{g}_{1,\grain} \right| & 0 & 0 \\
        0 & \left| \vec{g}_{2,\grain} \right| & 0 \\
        0 & 0 & \left| \vec{g}_{\sym,\grain}\right|
    \end{pmatrix}.
\end{align}

Then
\begin{align}
    \mathbf{UB} \cdot \mathbf{H}_\sym
    &=
    \mathbf{N}.
\end{align}

\item $\mathbf{U}$ can then be found by right-multiplying by $(\mathbf{BH}_\sym)^{-1}$.
\begin{align}
\mathbf{U} &= \mathbf{N} \cdot (\mathbf{BH}_\sym)^{-1}
\end{align}

\end{enumerate}

In the example above with $\vec{h}_\sym  = (1,\bar{1},0)^T$, $\vec{h}_1$ could be chosen to be $(1,1,0)^T$ and $\vec{h}_2$ could be $(0,0,1)^T$, resulting in
\begin{align}
    \mathbf{U} = \begin{pmatrix}
      \frac{1}{\sqrt{2}} & \frac{1}{\sqrt{2}} & 0 \\
      0 & 0 & 1 \\
      \frac{1}{\sqrt{2}} &  -\frac{1}{\sqrt{2}} & 0
    \end{pmatrix}.
\end{align}

\subsection{Proof that U is orthogonal}

Orthogonality requires that $\mathbf{U}^{-1} = \mathbf{U}^T$. $\mathbf{U}^T \cdot \mathbf{U} = \mathbf{1}$, or  equivalently $\left(\mathbf{U}^{-1}\right)^T \cdot \mathbf{U}^{-1} = \mathbf{1}$, which we show below.

\begin{align}
    \left(\mathbf{U}^{-1}\right)^T \cdot \mathbf{U}^{-1}
    &=
    \left( \mathbf{BH}_\sym \cdot \mathbf{N}^{-1} \right)^T
    \cdot
    \left( \mathbf{BH}_\sym \cdot \mathbf{N}^{-1} \right)
    \\
    &=
    \left(\mathbf{N}^{-1}\right)^T\cdot
    \mathbf{BH}_\sym^T  \cdot
    \cdot
    \mathbf{BH}_\sym \cdot \mathbf{N}^{-1}
    \\
    &=
    \mathbf{N}^{-1}
    \cdot
    \begin{pmatrix}
        \vec{g}_{1,\grain}^T \\
        \vec{g}_{2,\grain}^T \\
        \vec{g}_{\sym,\grain}^T
    \end{pmatrix}
    \cdot
    \begin{pmatrix}
        \vec{g}_{1,\grain}, \vec{g}_{2,\grain}, \vec{g}_{\sym,\grain}
    \end{pmatrix}
    \cdot
    \mathbf{N}^{-1}
    \\
    &=
    \mathbf{N}^{-1}
    \cdot
    \begin{pmatrix}
    \left| \vec{g}_{1,\grain} \right|^2 & 0 & 0 \\
    0 & \left| \vec{g}_{2,\grain} \right|^2 & 0 \\
    0 & 0 & \left| \vec{g}_{\sym,\grain} \right|^2
    \end{pmatrix}
    \cdot
    \mathbf{N}^{-1}
    \\
    &=
    \mathbf{N}^{-1}
    \cdot
    \mathbf{N}^{2}
    \cdot
    \mathbf{N}^{-1}
    \\
    &=
    \mathbf{1},
\end{align}
as, by construction, the vectors $\vec{q}_{(1,2,\sym),\grain}$ are mutually orthogonal.

\section{Example: Strain wave in the single crystal}
\label{sec:strain_wave}

For specific cases with enough restrictions on the components of the deformation gradient tensor, we demonstrate how a $\phi - 2\Delta\theta$ scan is enough to characterize the strain due to a longitudinal acoustic wave along a symmetric reflection. We choose the configuration as described in \cite{chalise2024formalism} with the case of a strain wave launched along the [100] direction of a single crystal diamond. This example is particularly useful in describing XFEL \citep{irvine2025dark} experiments visualizing the propagation of acoustic waves in materials. In \cite{chalise2024formalism}, the geometry of the experiment is such that $\mu = \theta_B$, $\omega = 0$, and $\eta = 0$. Since $\omega = 0$ and there are no rotational components of the deformation gradient introduced by the strain wave, we can set $\chi = 0$ to perform a strain scan with $\phi$ and $\Delta \theta$ as variables.

Since the strain wave is launched along the [100] direction, the only non-zero component of the strain resulting from the longitudinal strain wave is $\epsilon^{(g)}_{xx}$.

Therefore,
\begin{equation}
    \mathbf{F^{(g)}} = \begin{pmatrix} 1 + \epsilon^{(g)}_{xx} & 0 & 0 \\ 0 & 1 & 0 \\ 0 & 0 & 1 \end{pmatrix}
    \label{eq:strain_wave}
\end{equation}

Then,
\begin{equation}
    \mathbf{T} = \left(\mathbf{F^{(g)}}^T\right)^{-1} = \begin{pmatrix} \frac{1}{1 + \epsilon^{(g)}_{xx}} & 0 & 0 \\ 0 & 1 & 0 \\ 0 & 0 & 1 \end{pmatrix}
    \label{eq:strain_wave_T}
\end{equation}

\cite{chalise2024formalism} chooses the (400) diffraction plane for imaging. The $\mathbf{U}$ matrix that allows for the diffraction imaging in the given configuration is given by,
\begin{equation}
    \mathbf{U} = \begin{pmatrix} 0 & 0 & -1 \\ 0 & 1 & 0 \\ 1 & 0 & 0 \end{pmatrix}
    \label{eq:U_matrix_strain_wave}
\end{equation}

Since $\omega = \chi = 0$,
\begin{equation}
    \Gamma = \mathbf{R}_y(-\theta_B-\phi) = \mathbf{R}_y(-(\theta_B+\phi))
\end{equation}

and the left hand side of Eq.~\ref{eq:diffraction_condition_deformed_final} can be written as,
\begin{equation}
    \mathbf{R}_y(-(\theta_B+\phi)) \mathbf{U} \mathbf{T} \mathbf{B}_0 \begin{pmatrix} 4 \\ 0 \\ 0 \end{pmatrix}; \quad \text{where } \mathbf{B}_0 = \frac{2\pi}{a} \left( \begin{array}{ccc} 1 & 0 & 0 \\ 0 & 1 & 0 \\ 0 & 0 & 1 \end{array} \right)
\end{equation}
and a = 3.57 $\text{\AA}$ is the lattice parameter of the diamond. The left hand side further expands to,
\begin{equation}
    \begin{bmatrix}\cos(\theta_B + \phi) & 0 & -\sin(\theta_B + \phi) \\ 0 & 1 & 0 \\ \sin(\theta_B + \phi) & 0 & \cos(\theta_B + \phi) \end{bmatrix} \begin{bmatrix} 0 & 0 & -1 \\ 0 & 1 & 0 \\ 1 & 0 & 0 \end{bmatrix} \begin{bmatrix} \frac{1}{1 + \epsilon^{(g)}_{xx}} & 0 & 0 \\ 0 & 1 & 0 \\ 0 & 0 & 1 \end{bmatrix} \frac{2\pi}{a} \begin{bmatrix} 1 & 0 & 0 \\ 0 & 1 & 0 \\ 0 & 0 & 1 \end{bmatrix} \begin{pmatrix} 4 \\ 0 \\ 0 \end{pmatrix}
\end{equation}
and simplifies to,
\begin{equation}
    \frac{2\pi}{a} \cdot \frac{4}{1+\epsilon^{(g)}_{xx}} \begin{pmatrix} -\sin(\theta_B + \phi) \\ 0 \\ \cos(\theta_B + \phi) \end{pmatrix}.
\end{equation}

The right hand side of Eq.~\ref{eq:diffraction_condition_deformed_final} is given by,
\begin{equation}
    \frac{4\pi}{\lambda} \cdot \sin(\theta_B + \Delta \theta) \begin{pmatrix} -\sin(\theta_B + \Delta \theta) \\ 0 \\ \cos(\theta_B + \Delta \theta) \end{pmatrix}; \quad \text{since } \eta = 0.
\end{equation}

By equating the left hand side and the right hand side, we get,
\begin{align}
    \frac{2 \lambda}{a} \cdot \begin{pmatrix} \frac{-\sin(\theta_B + \phi)}{1+\epsilon^{(g)}_{xx}} \\ 0 \\ \frac{\cos(\theta_B + \phi)}{1+\epsilon^{(g)}_{xx}} \end{pmatrix} &= \sin(\theta_B + \Delta \theta) \begin{pmatrix} -\sin(\theta_B + \Delta \theta) \\ 0 \\ \cos(\theta_B + \Delta \theta) \end{pmatrix}.
    \label{eq:diffraction_condition_strain_wave}
\end{align}

From the Bragg's law for the (400) plane,
\begin{align}
    \frac{2 \lambda}{a} &= \sin(\theta_B)
\end{align}

Therefore, from Eq.~\ref{eq:diffraction_condition_strain_wave} we get,
\begin{equation}
    \frac{\sin(\theta_B)\cos(\theta_B + \phi)}{1+\epsilon^{(g)}_{xx}} = \sin(\theta_B + \Delta \theta) \cos(\theta_B + \Delta \theta)
    \label{eq:diffraction_condition_strain_wave_2}
\end{equation}

For small angles, the Eq.~\ref{eq:diffraction_condition_strain_wave_2} can be approximated as,
\begin{align}
    \frac{\theta_B(\theta_B + \phi)}{1+\epsilon^{(g)}_{xx}} &= (\theta_B + \Delta \theta)^2 
    \label{eq:diffraction_condition_strain_wave_1_approx}
\end{align}



Therefore the $\epsilon^{(g)}_{xx}$ is given by (for small deformations),
\begin{align}
    \epsilon^{(g)}_{xx} &= - \frac{\Delta \theta}{\theta_B}
\end{align}

From the above derivation, we can see that the $\epsilon^{(g)}_{xx}$ can be determined from a single reflection. And the $\epsilon^{(g)}_{xx}$ is the only non-zero component of the strain tensor $\mathbf{\varepsilon}$ for the strain wave launched along the [100] direction of the single crystal diamond. This is a special case where the $\mathbf{F^{(g)}}$ can be fully determined from a single reflection.

Now from Eq.~\ref{eq:diffraction_condition_strain_wave_1_approx} we get,
\begin{align}
    (\theta_B + \phi) \cdot \left( 1 + \frac{\Delta \theta}{\theta_B} \right) &= \left( 1 + \frac{\Delta \theta}{\theta_B} \right)^2 \\
    \phi &= \Delta \theta
\end{align}

This corresponds to the results described in \cite{chalise2024formalism} using the scalar form of Bragg's law.

\section{Absolute Sensitivity for \{202\} Reflections}
\label{absolute_sensitivity_202}

\begin{table}[H]
    \centering
    \begin{tabular}{|c|c|c|c|c|c|c|c|c|c|}
        \hline
        Parameter & $\phi_1$ & $\chi_1$ & $\Delta\theta_1$ & $\phi_2$ & $\chi_2$ & $\Delta\theta_2$ & $\phi_3$ & $\chi_3$ & $\Delta\theta_3$ \\
        \hline
        $E^{(g)}_{xx}$ & -0.304 & -0.587 & -0.116 & -0.678 & 0.852 & -0.116 & 0.000 & -0.000 & -0.000 \\
        $E^{(g)}_{yy}$ & -0.000 & 0.000 & 0.000 & 0.000 & 0.000 & 0.000 & 0.304 & 0.587 & -0.116 \\
        $E^{(g)}_{zz}$ & 0.695 & 0.826 & -0.116 & 0.321 & -0.562 & -0.116 & -0.695 & -0.826 & -0.116 \\
        $E^{(g)}_{xy}$ & -0.999 & 0.001 & 0.000 & -0.999 & -0.002 & 0.000 & 0.999 & -0.001 & 0.000 \\
        $E^{(g)}_{xz}$ & -0.393 & -0.240 & 0.232 & -0.356 & 0.290 & -0.231 & -0.999 & 0.001 & 0.000 \\
        $E^{(g)}_{yz}$ & 0.999 & -0.001 & 0.000 & -1.000 & 0.000 & 0.000 & 0.393 & 0.240 & 0.232 \\
        $R^{(g)}_{x}$ & 1.000 & 0.001 & 0.000 & -0.999 & -0.001 & 0.000 & -0.998 & -1.414 & 0.000 \\
        $R^{(g)}_{y}$ & -1.002 & -1.415 & 0.000 & 0.998 & -1.414 & 0.000 & 0.999 & -0.001 & 0.000 \\
        $R^{(g)}_{z}$ & 1.000 & 0.000 & 0.000 & 1.000 & -0.000 & 0.000 & 1.000 & -0.000 & 0.000 \\
        \hline
    \end{tabular}
    \caption{Absolute sensitivity for \{202\} reflections (rounded to three decimal points).}
    \label{tab:absolute_sensitivity_202}
\end{table}

\section{Angular Shift Analysis for \{400\} Reflections}
\label{Appendix4}

\begin{figure}[H]
    \centering
    \includegraphics[width=\linewidth]{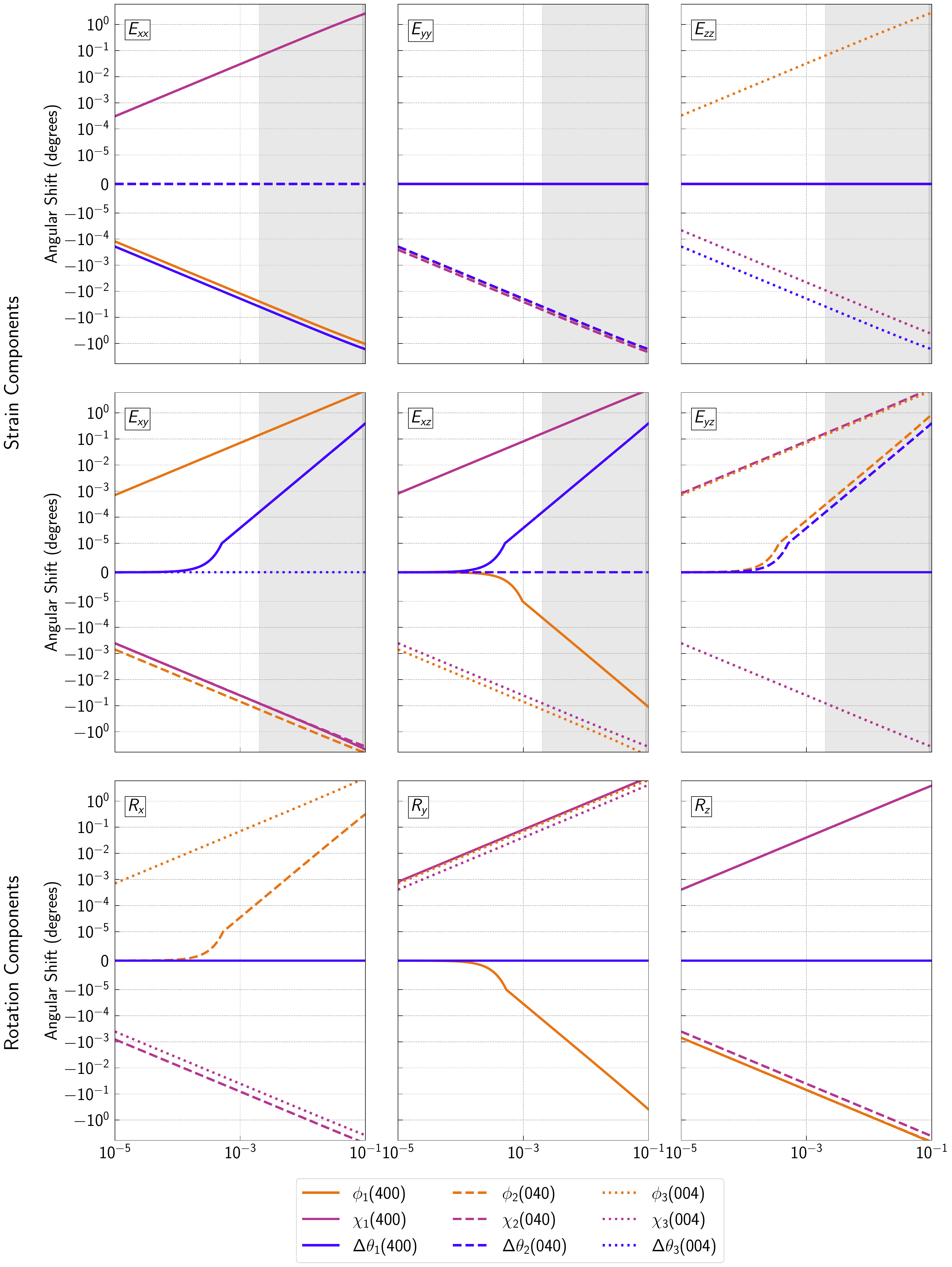}
    \caption{Angular shifts of $\phi$ $\chi$ and $\Delta\theta$ as a function of individual components of strain and lattice rotation for three symmetry-equivalent reflections ($(4, 0, 0)^T$, $(0, 4, 0)^T$, and $(0, 0, 4)^T$) in aluminum.}
    \label{fig:angularshifts311}
\end{figure}

We calculated the angular shifts for the three symmetry-equivalent reflections ($(4, 0, 0)^T$, $(0, 4, 0)^T$, and $(0, 0, 4)^T$) in aluminum to compare them with the results from Section~\ref{Deflection analysis}. In this simulation, the $h_{sym}$ is selected as $(1, 1, 1)^T$, with all other parameters unchanged as described in Section~\ref{Geometry used in the reconstruction}.

\section{Total Uncertainty in the Reconstruction}
\label{total_uncertainity}

Using the inequality relationship, Eq.~\ref{eq:absolute_sensitivity_linear} and Eq.~\ref{eq:error_in_angular_shifts} can be written as, 

\begin{align}
    \left\lVert \Delta \vec{\psi} \right\rVert \approx \left\lVert \boldsymbol{\nu} \cdot \Delta \vec{\varrho} \right\rVert &\leq \left\lVert \boldsymbol{\nu} \right\rVert \left\lVert \Delta \vec{\varrho} \right\rVert \\
    \frac{1}{\left\lVert\Delta \vec{\varrho}\right\rVert} &\leq \left\lVert \boldsymbol{\nu} \right\rVert \frac{1}{\left\lVert \Delta \vec{\psi} \right\rVert} \\
    \left\lVert{\delta}\vec{{\varrho}}\right\rVert \approx \left\lVert\boldsymbol{\nu}^{-1} \cdot \delta\vec{\psi} \right\rVert &\leq \left\lVert \boldsymbol{\nu}^{-1} \right\rVert \left\lVert \delta \vec{\psi} \right\rVert 
\end{align}

From these relationships we can write

\begin{equation}
    \frac{\left\lVert{\delta}\vec{{\varrho}}\right\rVert}{\left\lVert\Delta \vec{\varrho}\right\rVert} \leq \left\lVert \boldsymbol{\nu} \right\rVert \left\lVert \boldsymbol{\nu}^{-1} \right\rVert \frac{\left\lVert \delta \vec{\psi} \right\rVert}{\left\lVert \Delta \vec{\psi} \right\rVert}
\end{equation}

\section{Sobol Sensitivity Analysis}
\label{Appendix5}

Sobol sensitivity analysis decomposes the variance of angular shifts and separates contributions from individual deformation components and their higher-order interactions \citep{Sobolprime1993sensitivity}. To perform the Sobol sensitivity analysis, we chose strain values in the range of 10$^{-3}$ to 10$^{-1}$ and rotations in the range of 10$^{-3}$ to 10$^{-1}$~rad. We select this range to include possible effect of tensorial coupling of strain and rotations without the influence of sampling from the region that can be linearly approximated. The angular shifts $\mathbf{Y}$, which represent $\phi$, $\chi$, and $\Delta \theta$ for any reflection, serve as outputs. The nine components of $\mathbf{F^{(g)}}$ serve as inputs $X_i$ (where i = 1, 2, \ldots, 9 represents E$_{xx}$, E$_{yy}$, E$_{zz}$, E$_{xy}$, E$_{xz}$, E$_{yz}$, R$_x$, R$_y$, and R$_z$, respectively). The components of the strain and rotations are considered as statistically independent input parameters. The Sobol decomposition of the variance of the output $Y$ is expressed as
\begin{equation}
    \text{Var}(Y) = \sum_{i=1}^{9} \text{V}_{i} + \sum_{1 \leq i < j \leq 9} \text{V}_{ij} + \sum_{1 \leq i < j < k \leq 9} \text{V}_{ijk} + \ldots + \text{V}_{12\ldots9}.
\end{equation}
In this decomposition, $V_i$ represents the variance contribution from an individual deformation component $X_i$, while $V_{ij}$ represents the variance from interactions between components $X_i$ and $X_j$. $V_{12\ldots9}$ represents the highest-order interaction involving all nine deformation components simultaneously.

The first-order effect ($S_i$) measures the contribution of a single strain or rotation to the variance in angular shift normalized by the total variance. For example, $S_{E^{(g)}_{xx}}$ for $\phi$ (202) quantifies how much of the total variance in $\phi$ for the (202) reflection comes solely from the $E^{(g)}_{xx}$ strain component, excluding interactions with other components. The total sensitivity index ($S_{T_i}$) includes both the direct contribution (first-order) of the component of deformation  $X_i$ and its interactions with other components \citep{homma1996importance}. Because the total sensitivity contains the first-order contribution,  $0 \leq S_i \leq S_{T_i} \leq 1$. The second-order effect ($S_{ij}$) quantifies the contribution of the interaction between the primary component $X_i$ with the another components $X_j$. For instance, $S_{E^{(g)}_{xx},R^{(g)}_{y}}$ quantifies how the combined effect of $E^{(g)}_{xx}$ and $R^{(g)}_{y}$ on $\phi$ of (202) differs from the sum of their individual effects. 
\begin{figure}[H]
    \centering
    \includegraphics[width=\linewidth]{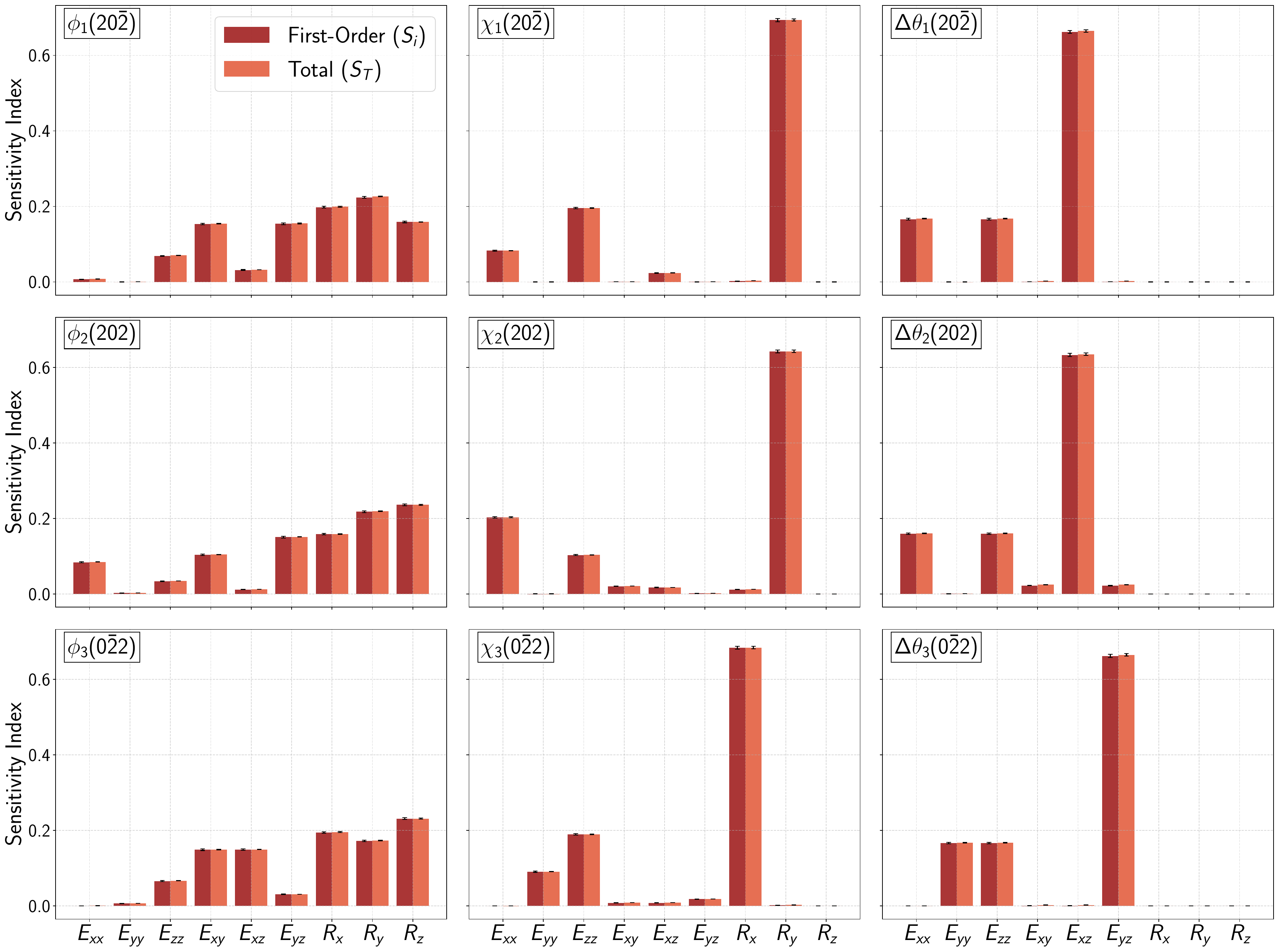}
    \caption{Sobol sensitivity analysis: Global sensitivities of strain and rotation components of the $\mathbf{F^{(g)}}$, sampled in between 10$^{-3}$ to 10$^{-1}$ for the strains and 10$^{-3}$ to 10$^{-1}$~rad for the rotations. Brown and dark orange bars represent the first order and total sensitivity of the individual strain and rotation component. The error bars show the statistical uncertainty in sensitivity indices across the parametric domain investigated in this study.}
    \label{fig:Sobol_sensitivity}
\end{figure}

Figure~\ref{fig:Sobol_sensitivity} shows the first-order and total sensitivity indices of strain and lattice rotations for the angular shifts. For this analysis, we used the Python library SALib \citep{herman2017salib}. We used the Saltelli sampler, a quasi-random sampling method, to generate 5,242,880 samples \citep{campolongo2011screening}.

Figure~\ref{fig:Sobol_sensitivity} shows trends identical to those discussed in the absolute sensitivity analysis for the given range of parameters. The sensitivities we observe in Figure~\ref{fig:absolute_sensitivity} are also well represented by the Sobol sensitivity analysis. Additionally, the trends observed in Figure~\ref{fig:deflection_anlysis} are well represented by the Sobol sensitivity analysis. Some angular shifts become more sensitive to large strains and rotations (e.g., $\chi$ for shear deformation), which are not evident in absolute sensitivity analysis. The Sobol method reflects these trends due to its ability to quantify the effects of interaction between strains and rotations. When users have limited knowledge of the expected magnitude of deformation, Sobol analysis helps to determine the optimal scan strategy.

\begin{table}[H]
    \centering
    \begin{tabular}{|c|c|c|c|c|c|}
        \hline
        Reflection & Sum $S_1$ & Sum $S_T$ & Sum Higher-Order & \% First-Order & \% Higher-Order \\
        \hline
        $\chi_{20\bar{2}}$ & 0.997904 & 1.002115 & 0.004211 & 99.58\% & 0.42\% \\
        $\chi_{202}$ & 0.998183 & 1.001829 & 0.003646 & 99.64\% & 0.36\% \\
        $\chi_{0\bar{2}2}$ & 0.998597 & 1.001411 & 0.002814 & 99.72\% & 0.28\% \\
        $\phi_{20\bar{2}}$ & 0.995754 & 1.004274 & 0.008520 & 99.15\% & 0.85\% \\
        $\phi_{202}$ & 0.997114 & 1.002904 & 0.005791 & 99.42\% & 0.58\% \\
        $\phi_{0\bar{2}2}$ & 0.997111 & 1.002899 & 0.005788 & 99.42\% & 0.58\% \\
        $\Delta\theta_{20\bar{2}}$ & 0.995433 & 1.004596 & 0.009163 & 99.09\% & 0.91\% \\
        $\Delta\theta_{202}$ & 0.995745 & 1.004283 & 0.008538 & 99.15\% & 0.85\% \\                                                         
        $\Delta\theta_{0\bar{2}2}$ & 0.995433 & 1.004596 & 0.009163 & 99.09\% & 0.91\% \\
        \hline
    \end{tabular}
    \caption{Sobol sensitivity analysis results for the angular shifts of the $(20\bar{2})$, $(202)$, and $(0\bar{2}2)$ reflections in aluminum. The table shows the sum of the first-order sensitivity indices ($S_1$), the sum of the total sensitivity indices ($S_T$), and the sum of the higher-order sensitivity indices ($S_{\text{Higher-Order}}$). The percentage of first-order effects is also provided for comparison.}
    \label{tab:Sobol_sensitivity_results}
\end{table}

Table~\ref{tab:Sobol_sensitivity_results} shows the sum of the first-order sensitivity indices ($S_1$), the sum of the total sensitivity indices ($S_T$), and the sum of the higher-order sensitivity indices ($S_{\text{Higher-Order}}$) for the angular shifts of the $(20\bar{2})$, $(202)$, and $(0\bar{2}2)$ reflections in aluminum. The percentage of first-order effects is also provided for comparison. For all angular shifts and reflections, the \% First-Order is over 99\%. This means that over 99\% of the variation in $\chi$, $\phi$, and $\Delta\theta$ is comes from to individual parameters (like $E^{(g)}_{xx}$ or $R^{(g)}_{x}$) acting independently. The \% Higher-Order is tiny, ranging from 0.28\% (for $\chi_{0\bar{2}2}$) to 0.91\% (for $\Delta\theta_{20\bar{2}}$). This suggests that interactions between parameters (e.g., $E^{(g)}_{xx}$ and $E^{(g)}_{xy}$ together) have a small impact. $\Delta\theta$ has slightly higher interaction effects (0.85\% to 0.91\%) compared to $\chi$ (0.28\% to 0.42\%) and $\phi$ (0.58\% to 0.85\%). This suggests $\Delta\theta$ is more sensitive to parameter interactions.

An important observation from the Sobol sensitivity analysis is that differences between total and first-order sensitivity indices are small. The maximum difference is 0.002614 for the $\Delta \theta$ $(20\bar{2})$ and  $E^{(g)}_{xz}$. This indicates, in our example, the variance is dominated by first-order effects.

When considering the $S_{\text{Higher-Order}}$ on the total sensitivity index for the angular shifts with large first order effects, we observe that the second and higher-order effects are not significant. For example, angular shift of $\chi$ due to $R^{(g)}_{y}$ for $(20\bar{2})$ and $R^{(g)}_{x}$ for $(0\bar{2}2)$ reflections are 0.0007 and 0.0004, respectively, with the interaction fraction ($\frac{(S_T - S_i)}{S_T}$) of 0.00098 and0.00036, respectively. However, for $\phi$, $\frac{(S_T - S_i)}{S_T}$ for $(20\bar{2})$ and $(0\bar{2}2)$ reflections is $\approx 1$, which means that the entire angular shift is contributed by the higher-order interactions. In this case, $(20\bar{2})$ and $(0\bar{2}2)$ did not have the components along $y$ and $x$ directions, respectively, to get influenced in the first-order. However, due to the tensorial coupling, they have non-zero sensitivities for large deformations. On the other hand, their total absolute sensitivity indices are still small: 0.00048 and 0.00047 for $(20\bar{2})$ and $(0\bar{2}2)$, respectively. The same behavior can be observed throughout the parametric domain where parameters with low magnitude of first-order sensitivity indices have high interaction fractions. Interactions are crucial for understanding the full influence of less dominant parameters. This behavior suggests that while some parameters show high interaction fractions, the absolute differences remain small, indicating that even dominant interactions may have limited effects on the total variance in angular shifts.

\end{myappendix}

\bibliography{bib}

\begin{thebibliography}{}

\bibitem[Abboud et~al., 2017]{Abboud2017}
Abboud, A., Kirchlechner, C., Keckes, J., {Conka Nurdan}, T., Send, S., Micha, J.~S., Ulrich, O., Hartmann, R., Str{\"{u}}der, L., and Pietsch, U. (2017).
\newblock {Single-shot full strain tensor determination with microbeam X-ray Laue diffraction and a two-dimensional energy-dispersive detector}.
\newblock {\em Journal of Applied Crystallography}, 50(3):901--908.

\bibitem[Ball et~al., 2024]{ball2024measurement}
Ball, O., Husband, R., McHardy, J., McMahon, M., Strohm, C., Kon{\^o}pkov{\'a}, Z., Appel, K., Cerantola, V., Coleman, A., Cynn, H., et~al. (2024).
\newblock Measurement bias in self-heating x-ray free electron laser experiments from diffraction studies of phase transformation in titanium.
\newblock {\em Journal of Applied Physics}, 136(11).

\bibitem[Barabash and Ice, 2014]{barabash2014strain}
Barabash, R.~I. and Ice, G. (2014).
\newblock {\em Strain and dislocation gradients from diffraction: spatially-resolved local structure and defects}.
\newblock World Scientific.

\bibitem[Bernier et~al., 2011]{bernier2011far}
Bernier, J.~V., Barton, N.~R., Lienert, U., and Miller, M.~P. (2011).
\newblock Far-field high-energy diffraction microscopy: a tool for intergranular orientation and strain analysis.
\newblock {\em The Journal of Strain Analysis for Engineering Design}, 46(7):527--547.

\bibitem[Bonet and Wood, 1997]{bonet1997nonlinear}
Bonet, J. and Wood, R.~D. (1997).
\newblock {\em Nonlinear continuum mechanics for finite element analysis}.
\newblock Cambridge university press.

\bibitem[Borgi et~al., 2024]{borgi2024simulations}
Borgi, S., R{\ae}der, T.~M., Carlsen, M.~A., Detlefs, C., Winther, G., and Poulsen, H.~F. (2024).
\newblock Simulations of dislocation contrast in dark-field x-ray microscopy.
\newblock {\em Journal of Applied Crystallography}, 57(2).

\bibitem[Busing and Levy, 1967]{busing1967angle}
Busing, W.~R. and Levy, H.~A. (1967).
\newblock Angle calculations for 3-and 4-circle x-ray and neutron diffractometers.
\newblock Technical report, Oak Ridge National Lab.(ORNL), Oak Ridge, TN (United States).

\bibitem[Callister and Rethwisch, 2022]{callister2022fundamentals}
Callister, W.~D. and Rethwisch, D.~G. (2022).
\newblock {\em Fundamentals of materials science and engineering}.
\newblock John Wiley \& Sons.

\bibitem[Campolongo et~al., 2011]{campolongo2011screening}
Campolongo, F., Saltelli, A., and Cariboni, J. (2011).
\newblock From screening to quantitative sensitivity analysis. a unified approach.
\newblock {\em Computer physics communications}, 182(4):978--988.

\bibitem[Cardoso and Leite, 2010]{cardoso2010exponentials}
Cardoso, J.~R. and Leite, F.~S. (2010).
\newblock Exponentials of skew-symmetric matrices and logarithms of orthogonal matrices.
\newblock {\em Journal of computational and applied mathematics}, 233(11):2867--2875.

\bibitem[Carlsen et~al., 2022]{carlsen2022fourier}
Carlsen, M., R{\ae}der, T.~M., Yildirim, C., Rodriguez-Lamas, R., Detlefs, C., and Simons, H. (2022).
\newblock Fourier ptychographic dark field x-ray microscopy.
\newblock {\em Optics Express}, 30(2):2949--2962.

\bibitem[Chalise and Cahill, 2023]{chalise2023electron}
Chalise, D. and Cahill, D.~G. (2023).
\newblock Electron paramagnetic resonance of n-type semiconductors for applications in three-dimensional thermometry.
\newblock {\em Physical Review Applied}, 20(6):064024.

\bibitem[Chalise et~al., 2022]{chalise2022temperature}
Chalise, D., Kenesei, P., Shastri, S.~D., and Cahill, D.~G. (2022).
\newblock Temperature mapping of stacked silicon dies from x-ray-diffraction intensities.
\newblock {\em Physical Review Applied}, 18(1):014076.

\bibitem[Chalise et~al., 2025]{chalise2024formalism}
Chalise, D., Wang, Y., Trigo, M., and Dresselhaus-Marais, L.~E. (2025).
\newblock Formalism to image the dynamics of coherent and incoherent phonons with dark-field x-ray microscopy using kinematic diffraction theory.
\newblock {\em Applied Crystallography}, 58(2).

\bibitem[Chen et~al., 2025]{chen20253d}
Chen, H., Hill, M.~O., Borgström, M.~T., and Wallentin, J. (2025).
\newblock 3d strain imaging of a heterostructured gainp/inp nanowire using bragg coherent diffraction x-ray imaging: Implications for optoelectronic devices.
\newblock {\em ACS Applied Nano Materials}.

\bibitem[Chung and Ice, 1999]{Chung1999}
Chung, J.-S. and Ice, G.~E. (1999).
\newblock {Automated indexing for texture and strain measurement with broad-bandpass x-ray microbeams}.
\newblock {\em Journal of Applied Physics}, 86(9):5249--5255.

\bibitem[Cretton et~al., 2025]{cretton2025observation}
Cretton, A. A.~W., Zelenika, A., Frankus, F., Y{\i}ld{\i}r{\i}m, C., Detlefs, C., Grumsen, F.~B., Winther, G., and Poulsen, H.~F. (2025).
\newblock Observation of the formation and sharpening of geometrically necessary boundaries.
\newblock {\em Materials Research Letters}, pages 1--9.

\bibitem[Detlefs et~al., 2025]{detlefs2025oblique}
Detlefs, C., Henningsson, A., Kanesalingam, B., Cretton, A., Corley-Wiciak, C., Frankus, F., Pal, D., Irvine, S., Borgi, S., Poulsen, H., et~al. (2025).
\newblock Oblique diffraction geometry for the observation of several non-coplanar bragg reflections under identical illumination.
\newblock {\em J. Appl. Cryst.}
\newblock in press.

\bibitem[Dresselhaus-Marais et~al., 2021]{dresselhaus2021situ}
Dresselhaus-Marais, L.~E., Winther, G., Howard, M., Gonzalez, A., Breckling, S.~R., Yildirim, C., Cook, P.~K., Kutsal, M., Simons, H., Detlefs, C., et~al. (2021).
\newblock In situ visualization of long-range defect interactions at the edge of melting.
\newblock {\em Science Advances}, 7(29):eabe8311.

\bibitem[He et~al., 2022]{he2022diamond}
He, Z., R{\"o}del, M., L{\"u}tgert, J., Bergermann, A., Bethkenhagen, M., Chekrygina, D., Cowan, T.~E., Descamps, A., French, M., Galtier, E., et~al. (2022).
\newblock Diamond formation kinetics in shock-compressed c- h- o samples recorded by small-angle x-ray scattering and x-ray diffraction.
\newblock {\em Science Advances}, 8(35):eabo0617.

\bibitem[Henningsson et~al., 2025]{henningsson2025towards}
Henningsson, A., Borgi, S., Winther, G., El-Azab, A., and Poulsen, H.~F. (2025).
\newblock Towards interfacing dark-field x-ray microscopy to dislocation dynamics modeling.
\newblock {\em arXiv preprint arXiv:2503.22022}.

\bibitem[Herman and Usher, 2017]{herman2017salib}
Herman, J. and Usher, W. (2017).
\newblock Salib: An open-source python library for sensitivity analysis.
\newblock {\em Journal of Open Source Software}, 2(9):97.

\bibitem[Hill et~al., 2022]{hill20223d}
Hill, M.~O., Schmiedeke, P., Huang, C., Maddali, S., Hu, X., Hruszkewycz, S.~O., Finley, J.~J., Koblmüller, G., and Lauhon, L.~J. (2022).
\newblock 3d bragg coherent diffraction imaging of extended nanowires: defect formation in highly strained ingaas quantum wells.
\newblock {\em ACS nano}, 16(12):20281--20293.

\bibitem[Hofmann et~al., 2017]{hofmann20173d}
Hofmann, F., Tarleton, E., Harder, R.~J., Phillips, N.~W., Ma, P.-W., Clark, J.~N., Robinson, I.~K., Abbey, B., Liu, W., and Beck, C.~E. (2017).
\newblock 3d lattice distortions and defect structures in ion-implanted nano-crystals.
\newblock {\em Scientific reports}, 7(1):45993.

\bibitem[Holzapfel, 2002]{holzapfel2002nonlinear}
Holzapfel, G.~A. (2002).
\newblock Nonlinear solid mechanics: a continuum approach for engineering science.

\bibitem[Homma and Saltelli, 1996]{homma1996importance}
Homma, T. and Saltelli, A. (1996).
\newblock Importance measures in global sensitivity analysis of nonlinear models.
\newblock {\em Reliability Engineering \& System Safety}, 52(1):1--17.

\bibitem[H{\"y}tch and Minor, 2014]{hytch2014observing}
H{\"y}tch, M.~J. and Minor, A.~M. (2014).
\newblock Observing and measuring strain in nanostructures and devices with transmission electron microscopy.
\newblock {\em MRS bulletin}, 39(2):138--146.

\bibitem[Irvine et~al., 2025]{irvine2025dark}
Irvine, S.~J., Katagiri, K., R{\ae}der, T.~M., Boesenberg, U., Chalise, D., Stanton, J.~I., Pal, D., Hallmann, J., Ansaldi, G., Brau{\ss}e, F., et~al. (2025).
\newblock Dark-field x-ray microscopy for 2d and 3d imaging of microstructural dynamics at the european x-ray free-electron laser.
\newblock {\em Journal of Applied Physics}, 137(5).

\bibitem[Isern et~al., 2025]{isern2024esrf}
Isern, H., Brochard, T., Dufrane, T., Brumund, P., Papillon, E., Scortani, D., Hino, R., Yildirim, C., Rodriguez~Lamas, R., Li, Y., Sarkis, M., and Detlefs, C. (2025).
\newblock The {ESRF} dark-field x-ray microscope at {ID03}.
\newblock {\em J. Phys. Conf. Ser.}, 3010:012163.

\bibitem[Jaisle et~al., 2023]{jaisle2023mhz}
Jaisle, N., C{\'e}bron, D., Kon{\^o}pkov{\'a}, Z., Husband, R.~J., Prescher, C., Cerantola, V., Dwivedi, A., Kaa, J.~M., Appel, K., Buakor, K., et~al. (2023).
\newblock Mhz free electron laser x-ray diffraction and modeling of pulsed laser heated diamond anvil cell.
\newblock {\em Journal of applied physics}, 134(9).

\bibitem[Katagiri et~al., 2023]{katagiri2023transonic}
Katagiri, K., Pikuz, T., Fang, L., Albertazzi, B., Egashira, S., Inubushi, Y., Kamimura, G., Kodama, R., Koenig, M., Kozioziemski, B., et~al. (2023).
\newblock Transonic dislocation propagation in diamond.
\newblock {\em Science}, 382(6666):69--72.

\bibitem[Lenhart et~al., 2002]{lenhart2002comparison}
Lenhart, T., Eckhardt, K., Fohrer, N., and Frede, H.-G. (2002).
\newblock Comparison of two different approaches of sensitivity analysis.
\newblock {\em Physics and Chemistry of the Earth, Parts A/B/C}, 27(9-10):645--654.

\bibitem[Lester and Aborn, 1925]{lester1925behavior}
Lester, H. and Aborn, R. (1925).
\newblock The behavior under stress of the iron crystals in steel: Part i.
\newblock {\em Army Ordnance}, 6(32):120--127.

\bibitem[Liesegang et~al., 2023]{liesegang2023investigation}
Liesegang, M., Lion, P., Beck, T., Gr{\"a}f, M., and Steidl, G. (2023).
\newblock Investigation of the tensile deformation behaviour in ni-based superalloy inconel alloy 617 using ebsd-based finite element simulations and optical flow method.
\newblock {\em Journal of Materials Science}, 58(21):8990--9005.

\bibitem[Liu et~al., 2004]{liu2004three}
Liu, W., Ice, G.~E., Larson, B.~C., Yang, W., Tischler, J.~Z., and Budai, J.~D. (2004).
\newblock The three-dimensional x-ray crystal microscope: A new tool for materials characterization.
\newblock {\em Metallurgical and Materials Transactions A}, 35:1963--1967.

\bibitem[Medu{\v{n}}a et~al., 2018]{meduvna2018lattice}
Medu{\v{n}}a, M., Isa, F., Jung, A., Marzegalli, A., Albani, M., Isella, G., Zweiacker, K., Miglio, L., and K{\"a}nel, H.~v. (2018).
\newblock Lattice tilt and strain mapped by x-ray scanning nanodiffraction in compositionally graded sige/si microcrystals.
\newblock {\em Applied Crystallography}, 51(2):368--385.

\bibitem[Nie et~al., 2019]{nie2019approaching}
Nie, A., Bu, Y., Li, P., Zhang, Y., Jin, T., Liu, J., Su, Z., Wang, Y., He, J., Liu, Z., et~al. (2019).
\newblock Approaching diamond’s theoretical elasticity and strength limits.
\newblock {\em Nature communications}, 10(1):5533.

\bibitem[Pal et~al., 2025]{pal2025measuring}
Pal, D., Wang, Y., Gurunathan, R., and Dresselhaus-Marais, L. (2025).
\newblock Measuring the burgers vector of dislocations with dark-field x-ray microscopy.
\newblock {\em Applied Crystallography}, 58(1).

\bibitem[Pateras et~al., 2020]{pateras2020combining}
Pateras, A., Harder, R., Cha, W., Gigax, J.~G., Baldwin, J.~K., Tischler, J., Xu, R., Liu, W., Erdmann, M.~J., Kalt, R., et~al. (2020).
\newblock Combining laue diffraction with bragg coherent diffraction imaging at 34-id-c.
\newblock {\em Synchrotron Radiation}, 27(5):1430--1437.

\bibitem[Poulsen et~al., 2021]{poulsen2021geometrical}
Poulsen, H., Dresselhaus-Marais, L., Carlsen, M., Detlefs, C., and Winther, G. (2021).
\newblock Geometrical-optics formalism to model contrast in dark-field x-ray microscopy.
\newblock {\em Journal of Applied Crystallography}, 54(6):1555--1571.

\bibitem[Poulsen et~al., 2017]{poulsen2017x}
Poulsen, H.~F., Jakobsen, A., Simons, H., Ahl, S.~R., Cook, P., and Detlefs, C. (2017).
\newblock X-ray diffraction microscopy based on refractive optics.
\newblock {\em Journal of Applied Crystallography}, 50(5):1441--1456.

\bibitem[Schlenker et~al., 1978]{Schlenker1978}
Schlenker, J.~L., Gibbs, G.~V., and Boisen~Jnr, M.~B. (1978).
\newblock Strain-tensor components expressed in terms of lattice parameters.
\newblock {\em Acta Crystallographica Section A}, 34(1):52--54.

\bibitem[{\c{S}}eren et~al., 2023]{cseren2023representative}
{\c{S}}eren, M.~H., Pagan, D.~C., and Noyan, I.~C. (2023).
\newblock Representative volume elements of strain/stress fields measured by diffraction techniques.
\newblock {\em Applied Crystallography}, 56(4):1144--1167.

\bibitem[Simons et~al., 2017]{simons2017simulating}
Simons, H., Ahl, S.~R., Poulsen, H.~F., and Detlefs, C. (2017).
\newblock Simulating and optimizing compound refractive lens-based x-ray microscopes.
\newblock {\em Journal of synchrotron radiation}, 24(2):392--401.

\bibitem[Simons et~al., 2015]{simons2015dark}
Simons, H., King, A., Ludwig, W., Detlefs, C., Pantleon, W., Schmidt, S., St{\"o}hr, F., Snigireva, I., Snigirev, A., and Poulsen, H.~F. (2015).
\newblock Dark-field x-ray microscopy for multiscale structural characterization.
\newblock {\em Nature communications}, 6(1):6098.

\bibitem[Sobolprime, 1993]{Sobolprime1993sensitivity}
Sobolprime, I. (1993).
\newblock Sensitivity analysis for nonlinear mathematical models.
\newblock {\em Math. Model. Comput. Exp}, 1:407--414.

\bibitem[Tanner, 2013]{tanner2013x}
Tanner, B.~K. (2013).
\newblock {\em X-ray diffraction topography: international series in the science of the solid state}, volume~10.
\newblock Elsevier.

\bibitem[Tanner et~al., 2021]{tanner2021x}
Tanner, B.~K., McNally, P.~J., and Danilewsky, A.~N. (2021).
\newblock X-ray imaging of silicon die within fully packaged semiconductor devices.
\newblock {\em Powder Diffraction}, 36(2):78--84.

\bibitem[Truesdell et~al., 2004]{truesdell2004non}
Truesdell, C., Noll, W., Truesdell, C., and Noll, W. (2004).
\newblock {\em The non-linear field theories of mechanics}.
\newblock Springer.

\bibitem[Wang et~al., 2025]{wang2025computing}
Wang, Y., Bertin, N., Pal, D., Irvine, S.~J., Katagiri, K., Rudd, R.~E., and Dresselhaus-Marais, L.~E. (2025).
\newblock Computing virtual dark-field x-ray microscopy images of complex discrete dislocation structures from large-scale molecular dynamics simulations.
\newblock {\em Applied Crystallography}, 58(2).

\bibitem[Wilkinson et~al., 2006]{wilkinson2006high}
Wilkinson, A.~J., Meaden, G., and Dingley, D.~J. (2006).
\newblock High-resolution elastic strain measurement from electron backscatter diffraction patterns: New levels of sensitivity.
\newblock {\em Ultramicroscopy}, 106(4-5):307--313.

\bibitem[Yildirim et~al., 2024]{yildirim2024understanding}
Yildirim, C., Flatscher, F., Ganschow, S., Lassnig, A., Gammer, C., Todt, J., Keckes, J., and Rettenwander, D. (2024).
\newblock Understanding the origin of lithium dendrite branching in li6. 5la3zr1. 5ta0. 5o12 solid-state electrolyte via microscopy measurements.
\newblock {\em Nature Communications}, 15(1):8207.

\bibitem[Yildirim et~al., 2023]{yildirim2023extensive}
Yildirim, C., Poulsen, H.~F., Winther, G., Detlefs, C., Huang, P.~H., and Dresselhaus-Marais, L.~E. (2023).
\newblock Extensive 3d mapping of dislocation structures in bulk aluminum.
\newblock {\em Scientific Reports}, 13(1):3834.

\bibitem[Zelenika et~al., 2025]{zelenika2025observing}
Zelenika, A., Cretton, A. A.~W., Frankus, F., Borgi, S., Grumsen, F.~B., Yildirim, C., Detlefs, C., Winther, G., and Poulsen, H.~F. (2025).
\newblock Observing formation and evolution of dislocation cells during plastic deformation.
\newblock {\em Scientific Reports}, 15(1):8655.

\end{thebibliography}

\end{document}